# Towards responsible AI for education: Hybrid human-AI to confront the Elephant in the room

Danial Hooshyar[1,2], Gustav Šír[3], Yeongwook Yang[4], Eve Kikas[5], Raija Hämäläinen[2], Tommi Kärkkäinen[6], Dragan Gašević[7], Roger Azevedo[8]

[1]School of digital Technologies, Tallinn University, Tallinn, Estonia
[2]Department of Education, University of Jyväskylä, Jyväskylä, Finland
[3]Department of Computer Science, Czech Technical University, Prague, Czech Republic
[4]Department of Computer Science and Engineering, Gangneung-Wonju National University
Wonju, Republic of Korea
[5]School of Natural Sciences and Health, Tallinn University, Tallinn, Estonia
[6]Faculty of Information Technology, University of Jyväskylä, Jyväskylä, Finland
[7]Faculty of Information Technology, Monash University, Clayton, Victoria, Australia
[8]School of Modeling Simulation and Training, University of Central Florida, Orlando, USA

Despite significant advancements in AI-driven educational systems and ongoing calls for responsible AI for education, multiple critical issues remain unresolved—acting as the *elephant in the room* within the AI in education, learning analytics, educational data mining, learning sciences, and educational psychology communities. This critical analysis paper identifies and examines nine persistent challenges that continue to undermine the fairness, transparency, and effectiveness of current AI methods and applications in education. These include: 1) the lack of clarity around what AI for education truly means—often ignoring the distinct purposes, strengths, and limitations of different AI families—and the growing trend of equating it with domain-agnostic, company-driven large language models; 2) the widespread neglect of essential learning processes such as motivation, emotion, and (meta)cognition in AI-driven learner modelling and their contextual nature; 3) the limited integration of domain knowledge and lack of stakeholder involvement in AI design and development; 4) the continued use of non-sequential machine learning models on temporal educational data, which fails to capture the dynamics and dependencies critical to understanding real learning progressions; 5) the misuse of non-sequential metrics to evaluate sequential or temporal models; 6) using unreliable explainable AI methods to provide explanations for black-box models; 7) ignoring ethical guidelines in addressing data inconsistencies during model training; 8) use of mainstream AI methods for pattern discovery and learning analytics without systematic benchmarking, which risks leading to unreliable conclusions; and 9) focusing on global prescriptions while overlooking localized, student-specific recommendations. Supported by both theoretical and empirical research, we demonstrate how hybrid AI methods—specifically neural-symbolic AI—address the *elephant in the room* and serve as the foundation for building responsible, trustworthy AI systems that align with the needs of students, teachers, and policymakers.

## 1. Introduction

The rapid adoption of artificial intelligence (AI) in education has introduced powerful tools to support teaching and learning (Azevedo & Wiedbusch, 2023; Vincent-Lancrin & Van der Vlies, 2020). Governments worldwide have begun employing AI for education on a large scale and are institutionalizing its use. One prominent example is Estonia's recent AI Leap program (https://tihupe.ee/en), which mandates the integration of AI tools in schools to support education. While such initiatives are promising, they also demand careful scrutiny. The effectiveness of AI for education is not just about adoption but about ensuring responsible and trustworthy implementation (European Union, 2024). This requires clearly defining what kind of AI is being used, for what purpose, and with what ethical safeguards, as their absence can lead to serious consequences. A prominent example is the United Kingdom's unsuccessful attempt to use AI—specifically the A-level algorithm—to automate student grading, where the estimated grades were systematically biased against certain students and schools (Heaton et al., 2023). Such concerns have been highlighted by recent calls from organizations like UNESCO (UNESCO, 2019) and regulatory safeguards like the EU AI Act (European Union, 2024; Goodman & Flaxman, 2017), urging a move towards responsible and trustworthy AI, particularly in high-stakes domains like education, where black-box AI models can reinforce inequalities. To this end, Rudin (2019) advocates against using such opaque systems in critical settings when interpretable alternatives with comparable performance exist. Relying on inscrutable AI models in education, therefore, poses not only ethical (Durán & Jongsma, 2021; Lo Piano, 2020) but also legal (European Union, 2024) challenges. Considering the recent classification of education as a high-risk domain for AI applications, trustworthiness and interpretability emerge as fundamental ethical imperatives, not merely technical objectives.

Recently, researchers have articulated the notion of *responsible AI* in different ways, each grounded in principles such as fairness, privacy, accountability, and transparency. Maree et al. (2020), for instance, emphasize these core principles along with soundness, while Arrieta et al. (2020) broaden the discussion to encompass ethics, safety, and security Additional viewpoints, including those of Eitel-Porter (2020) and Werder et al. (2022), underscore the essential role of explainability. A more comprehensive perspective is presented by Jakesch et al. (2022), who advocate for responsible AI that supports human dignity, inclusivity, societal benefit, sustainability, and autonomy. Drawing from this evolving body of work, Goellner et al. (2024) propose a comprehensive yet practical definition: "*Responsible AI is human-centred and ensures users' trust through ethical ways of decision making. The decision-making must be fair, accountable, not biased, with good intentions, non-discriminating, and consistent with societal laws and norms. Responsible AI ensures that automated decisions are explainable to users while always preserving users' privacy through a secure implementation*." The urgency of responsible AI adoption in education is further underscored by multiple research studies that have addressed the key principles of responsible AI for education (Fu & Weng, 2024; Pargman et al., 2024; Porayska-Pomsta, 2024). While they highlight the importance of transparency, accountability, and human-centred design and development in AI-driven educational technologies, much of the discussion remains theoretical, with limited implementation of such hybrid human-centred AI methods that are trustworthy, ethical, and interpretable.

Responsible AI use requires responsible development. Without it, enforcing ethical use alone cannot correct foundational flaws—as exemplified by large language models (Resnik, 2024). Ensuring responsibility throughout the entire AI lifecycle—starting from development—prevents systemic issues that could otherwise compromise ethical, trustworthy, and responsible deployment. According to the Director of the Department of Education at the University of Oxford[1], *"There's no way you can improve on the data when that's all you have available. These systems only reproduce what's gone before and entrench inequalities already in the data, which is a real problem. So, you have to be careful of the ethics of how this is done and move very slowly. Educational technology sales pitches make big claims that don't pan out, so technological development is needed, but the real issue is conceptual. It's about understanding what the models should look like and what the data can and can't do—and this will only get more important as we move to AI."*

Hybrid human-AI methods have shown great potential for achieving responsible AI for education through the combination of expert knowledge and data-driven approaches, helping to ensure that AI systems deliver not just accuracy, but also transparency and accountability. Among the most widely used methods are Bayesian networks, valued for their ability to combine expert knowledge with machine learning (Pearl, 1988). Their interpretability and latent variable modelling make them particularly useful for tracking students' evolving learning processes (Almond et al., 2015; Hooshyar & Druzdzel, 2024; Weidlich et al., 2022). However, Bayesian networks face significant challenges, including high computational complexity and dependence on expert-defined structures (Daly et al., 2011). Conversely, efficient, purely data-driven methods like deep neural networks surpass Bayesian networks in predictive performance in handling both sequential and non-sequential learning data. By automating feature extraction, deep neural networks eliminate the need for manual input, making them highly scalable and adaptable (Dhar, 2024; Gervet et al., 2020; Hooshyar et al., 2022a; Y. Mao, 2018). Despite their advantages, such purely data-driven models often lack interpretability and fail to incorporate domain knowledge, resulting in opaque decision-making and potential biases, which have driven the field towards explainable AI (XAI) concepts in pursuit of responsible AI (Khosravi et al., 2022).

Recent research has largely focused on enhancing the interpretability of data-driven methods, yet often ignore explicitly embedding domain knowledge. To improve the transparency of purely data-driven methods in education, many researchers have explored post-hoc XAI algorithms such as SHAP (Lundberg & Lee, 2017) and LIME (Ribeiro et al., 2016) including (Hooshyar et al., 2022a; Saarela et al., 2021). Despite their popularity, these methods have notable limitations. A key issue is their tendency to base explanations on particular input features used in predictions, without systematically incorporating these features into the explanation itself—an omission that can lead to misleading insights (Lakkaraju & Bastani, 2020; Slack et al., 2020). Additionally, some XAI methods have shifted focus away from fidelity, which refers to how well an explanation mirrors the actual behaviour of the underlying model. As a result, the generated explanations may not consistently represent the model's true decision-making process (Hooshyar & Yang, 2024b; Rudin, 2019; White & d'Avila Garcez, 2020).

---
[1] https://diginomica.com/uk-level-algorithm-fiasco-global-example-what-not-do-what-went-wrong-and-why

Another powerful hybrid human-AI approach, neural-symbolic AI[2] (NSAI)—often called the "third wave of AI"—has emerged as a promising method that integrates symbolic reasoning with neural networks to enhance interpretability and robustness (Garcez & Lamb, 2023; Hooshyar & Yang, 2021). NSAI combines the structured reasoning of symbolic models with the pattern recognition capabilities of deep learning. Through the incorporation of symbolic knowledge into neural networks (Sourek et al., 2018), NSAI helps align AI outputs with established educational frameworks, enhancing both generalizability and interpretability (Hooshyar et al., 2024; Tato & Nkambou, 2022).

Despite the above-mentioned challenges and the potential of hybrid human-AI methods like neural-symbolic computing to address them—an important step towards responsible and trustworthy AI for education—several critical issues remain unresolved, acting as the *elephant in the room* within the AI in education, learning analytics, educational data mining, learning sciences, and educational psychology communities. These challenges, primarily technical, often go unaddressed or are acknowledged but overlooked, contributing to the low uptake of AI tools in practice. This position paper highlights these challenges and proposes hybrid human-AI approaches, such as NSAI, as a pathway towards more transparent, fair, and responsible AI development, facilitating the responsible use of AI for education. Supported by both theoretical and empirical research, we demonstrate how hybrid AI methods address key challenges in AI for education and serve as the foundation for building responsible, trustworthy AI systems that align with the needs of students, teachers, and policymakers.

## 2. The paradigm shifts in AI for education: From expert systems to general intelligence

The field of AI has undergone several paradigm shifts, each addressing significant limitations of its predecessor (Du Boulay et al., 2023). Fig. 1 illustrates the overall paradigm shifts in AI – inspired by Dhar (2024), showing the evolution from *expert systems* that rely on highly curated expert knowledge and rules, to *machine learning*, which discovers patterns from structured datasets, and *deep learning*, which learns latent representations from loosely curated data. The shift progresses towards *general intelligence*, which aims to understand the world from minimally curated real-world data. Each stage addresses key bottlenecks: knowledge engineering in expert systems, feature engineering in machine learning, and customization in deep learning. These shifts, driven by advancements in data availability, computational power, and algorithmic improvements, have transformed AI from rule-based reasoning to complex statistical pattern recognition (Dhar, 2024).

---

[2] Researchers have commonly referred to the integration of neural and symbolic artificial intelligence as "neural-symbolic," though the term "neurosymbolic" is also frequently used in both scholarly publications and the media.

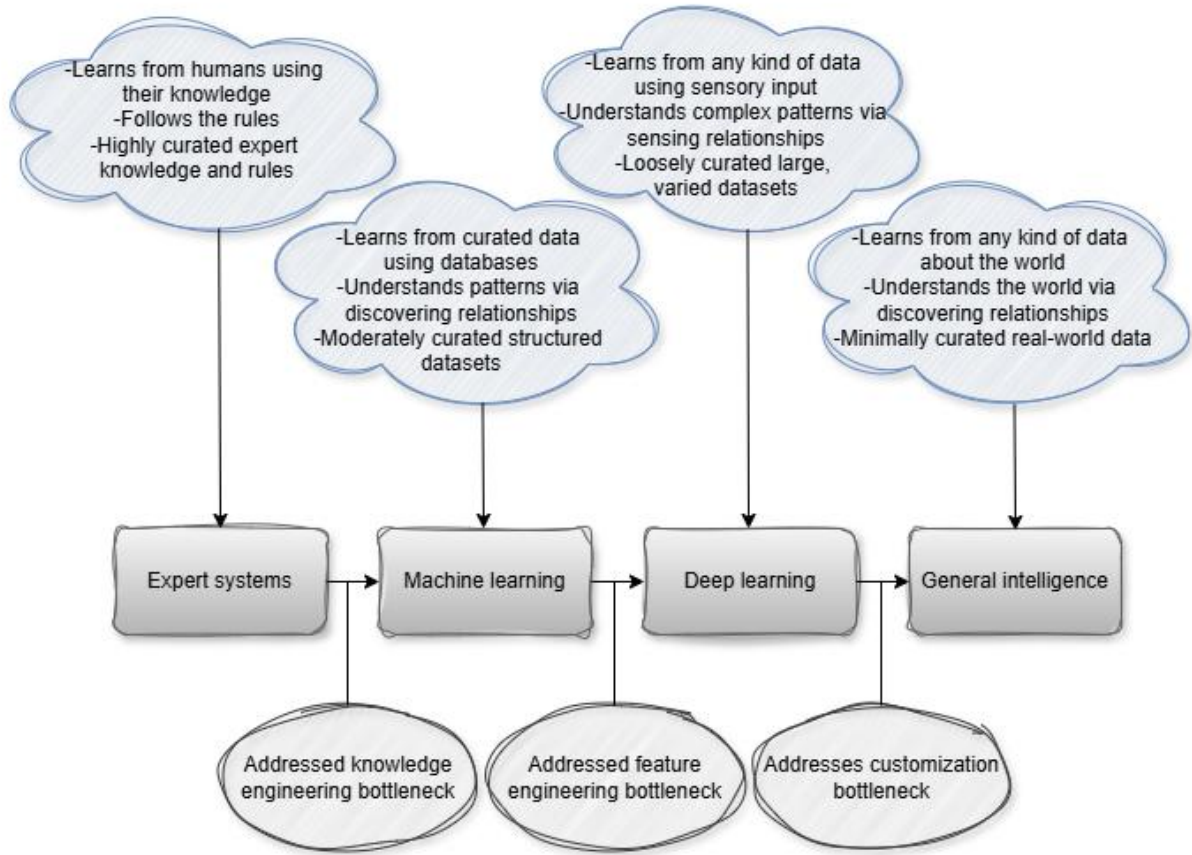

**Fig. 1** Overall paradigm shift of artificial intelligence – inspired by Dhar (2024).

*Expert systems* were a dominant approach in AI research from the 1960s to the late 1980s, designed to replicate human decision-making. These systems relied on rule bases, typically structured as IF/THEN rules, to draw inferences. Early successes included intelligent tutoring systems such as GUIDON, which used expert-defined pedagogical strategies to guide students through case-based learning in structured domains (Clancey, 1979). One of the key strengths of expert systems is their transparent, step-by-step reasoning, which enhances control, explainability, and troubleshooting—critical factors that are especially important in education, where transparency and trust are crucial (Conati et al., 2018a). Furthermore, rule-based approaches are modular in nature, enabling updates to specific rules without impacting the whole system, making them well-suited for handling structured educational tasks (Hatzilygeroudis & Prentzas, 2004). Nonetheless, expert systems also had significant limitations. The most notable was the knowledge engineering bottleneck—the challenge of extracting, encoding, and maintaining human expertise in rule-based formats (Lenat et al., 1985). Moreover, these systems struggled to handle uncertainty and lacked the adaptability required to go beyond predefined rules (McDermott, 1982). These constraints ultimately contributed to their decline as more flexible, data-driven AI approaches emerged.

*Machine learning* (ML) gained prominence in the late 1980s and 1990s as AI shifted away from manually encoded rules towards data-driven discovery of statistical relationships. Unlike expert systems that required humans to define relationships explicitly, ML enabled models to learn these relationships autonomously from large datasets through optimization and statistical inference (Weiss & Kulikowski, 1991). This transition was driven by advances in database technology, the rise of the Internet, and the explosion of available digital data. Instead of hardcoded IF/THEN rules, ML algorithms optimize a loss function to minimize prediction errors, making AI more adaptive and capable of generating knowledge from empirical data rather than just applying predefined logic (Breiman et al., 1984). Self (1986) and Webb (1989) discussed the application of machine learning to student modelling, while Webb et al. (1991) demonstrated its potential in intelligent tutoring systems. However, this approach introduced a new challenge: feature engineering. Humans still had to manually extract meaningful features from raw data, such as identifying key attributes in educational assessments or learning behaviours in digital environments (Dhar, 2024). One recent example include Hooshyar et al. (2022a), who introduced StudModel, an educational data mining approach that predicts students' risk of failure by analysing content access, engagement, and assessment behaviour, relying on feature engineering to construct these three core dimensions. This bottleneck led to the rise of deep learning, which sought to bypass manual feature engineering by allowing AI models to process raw inputs directly, further advancing automation (LeCun & Bengio, 1995).

*Deep learning* emerged as a transformative AI paradigm by overcoming the feature-engineering bottleneck, enabling models to automatically discover hierarchical patterns and learn directly from raw sensory data such as images, text, and audio (Hinton, 1992; LeCun & Bengio, 1995). The architecture of deep neural networks (DNNs) consists of multiple stacked layers of artificial neurons, where each successive layer refines and abstracts the input data, translating raw sensory input into useful learning representations. Advances in hardware made large-scale training feasible, allowing deep learning to scale to complex real-world applications (Hooshyar & Yang, 2024a; Krizhevsky et al., 2012). A key strength of DNNs is their ability of universal function approximation, allowing modelling intricate patterns in high-dimensional data without explicit programming (Aggarwal, 2018). Some examples include the use of sequential deep neural networks for the knowledge tracing task, where deep learning could enhance prediction performance and capture complex student learning patterns (Gervet et al., 2020; Piech et al., 2015), as well as image-based learner modelling, where convolutional neural networks (LeCun & Bengio, 1995) and transfer learning have been used to predict learners' performance in computational thinking (Hooshyar & Yang, 2024a). However, this shift introduced challenges in learning biases and interpretability. Unlike expert systems that encode domain knowledge for transparent and controllable decision-making, deep neural networks distribute knowledge across weighted connections, making their reasoning difficult to trace. This could lead to reliance on spurious correlations, reducing generalizability and potentially reinforcing biases in educational decision-making (Hooshyar et al., 2024). The lack of interpretability in these models makes it difficult to foster trust and ensure fairness in AI-driven decisions (Hooshyar & Yang, 2024b).

*General intelligence* in AI has evolved with pre-trained models, integrating both general and specialized knowledge, surpassing task-specific optimization. Unlike earlier AI applications that struggled with transferability, general intelligence enables models to apply learned knowledge across diverse domains and unforeseen scenarios (Dhar, 2024). Each paradigm shift—from expert systems structuring knowledge, to machine learning utilizing structured databases, to deep learning processing unstructured real-world data—has brought AI closer to human-like adaptability. The rise of pre-trained models departs from earlier rule-based approaches by embedding common sense and experiential knowledge through large-scale self-supervised learning (Hoffmann et al., 2022). This shift has been driven by the vast expansion of data sources, including language, images, and social interactions, allowing AI to develop fluency in natural communication. By overcoming the bottlenecks of previous paradigms, AI can now integrate expertise, common sense, and tacit knowledge, enhancing its ability to reason across different contexts. Scaling laws indicate that increasing model complexity, data size, and computational power continue to improve AI performance, with no clear limits yet identified. Additionally, advancements in multi-modal learning, where AI integrates sensory inputs such as vision, sound, and touch, mirror human perception and further push the boundaries of general intelligence. However, while scaling has fuelled remarkable progress, fundamental aspects of intelligence remain elusive and unlikely to be solved merely by increasing model size and complexity (Kasneci et al., 2023; Resnik, 2024).

The current AI paradigm, while transformative, is not necessarily the ultimate or most effective one in all respects. Like previous paradigm shifts in science, it has introduced remarkable advancements, such as deep learning's unparalleled capabilities in perception, but it remains fundamentally flawed in areas such as explainability and transparency (Conati et al., 2018a; Dhar, 2024). The opacity of AI models—particularly those built on deep learning—raises serious concerns about trust, as these systems often provide predictions without revealing the underlying reasoning. This is evident in high-stakes applications, where AI can offer valuable insights yet struggles to explain its decision-making process (Hooshyar et al., 2024; Hooshyar & Yang, 2024b; Khosravi et al., 2022). Additionally, large language models (LLMs) in education can generate responses that seem credible but may sometimes be fabricated or biased due to the uncurated nature of their training data (Kasneci et al., 2023). The unpredictability of AI systems, including their tendency to produce inconsistent outputs to the same input, further undermines confidence in their reliability. While efforts like reinforcement learning via human feedback attempt to impose safeguards, these measures do not always succeed in preventing undesirable behaviours, and the AI's inner workings remain largely inscrutable (Dhar, 2024; Havrilla et al., 2024). Moreover, this opacity allows biases present in training data to persist unchecked, as there is no reliable mechanism to distinguish between harmless patterns and harmful generalizations. As argued by Resnik (2024), LLMs are fundamentally shaped by the statistical properties of language rather than an intrinsic understanding of meaning or ethics. Their representations are built purely on distributional patterns, making it difficult to separate factual knowledge from societal biases. This issue is exemplified by how an LLM might learn that a nurse is a healthcare worker (a factual definition), that nurses often wear blue (a contingent but normatively acceptable fact), and that nurses are more often associated with female pronouns (a contingent but problematic societal bias). Because LLMs lack an innate mechanism to distinguish between these categories, biases remain deeply embedded in their learned representations, reinforcing existing stereotypes in ways that are difficult to control or mitigate. This becomes especially concerning in critical areas such as healthcare and education, where AI-driven decisions can significantly impact individuals' lives. In such contexts, a more responsible AI paradigm is essential—one that prioritizes fairness, transparency, and accountability to ensure AI serves society's best interests.

### 3. Challenges of current AI methods in education: The Elephant in the room

This section explores (technical) challenges and limitations of existing AI methods in education that hinder their potential and contribute to distrust among key stakeholders, including teachers, learning science researchers, parents, and students. These challenges have led to the slow adoption of AI in schools. Additionally, the section highlights overlooked areas where AI can be improved to enhance its practicality and effectiveness in real-world educational settings.

### 3.1. What do we really mean by AI for Education?

As explained in the *paradigm shift* section (Section 2), each family of AI has its own advantages, disadvantages, and specific purpose, yet they are often overlooked and treated as a single entity (L. Chen et al., 2020; Forero-Corba & Bennasar, 2024; Mouta et al., 2024; Wang et al., 2024). For instance, in their review of AI for education, Wang et al. (2024) and Chen et al. (2020) did not include expert systems, while Forero-Corba and Bennasar (2024) focused solely on machine learning, illustrating the selective scope and fragmented coverage of AI methodologies in education research. This issue has been exacerbated by the rise of LLMs. Generative AI (GenAI) methods—often mistakenly equated only with LLMs—have dominated the discourse in the education community, despite their earlier and broader applications in synthetic data augmentation using variational autoencoders (Giannakos et al., 2024; Kingma & Welling, 2019; Yan, Greiff, et al., 2024) or generative adversarial networks (Goodfellow et al., 2020). As a result, AI for education is increasingly being framed around LLMs, with less attention to the purpose, limitations, and strengths of different AI methods. To investigate this trend, we conducted a search in the Web of Science using the query TS = (("artificial intelligence" OR "AI") AND (education)), which returned 17,088 articles, with 6,772 published in 2024 alone (see Fig. 2), where the community began a strong shift towards integrating LLMs in education without considering their continuously proven weaknesses (İpek et al., 2023; Lo, 2023). To refine our focus, we further searched for TS=((education) AND ("large language model*" OR LLM OR genAI OR "generative AI" OR chatgpt)), retrieving 4,551 articles, with 3,009 from 2024 (see Fig. 3), indicating the growing dominance of LLMs in educational research. While a more rigorous screening process, following standard methodologies such as PRISMA (Page et al., 2021), is necessary to draw definitive conclusions, these preliminary findings underscore the increasing popularity and adoption of LLM-based AI for education—often without sufficient critical evaluation of its limitations. Among these studies, many highlight the benefits of LLMs in education, such as improving student learning performance and engagement in computer programming (H. Kumar et al., 2023; Lyu et al., 2024) and enhancing teaching assistant efficiency (Miroyan et al., 2025), among others. However, a growing number of articles underline the biases and weaknesses of this AI family (J. Lee et al., 2024; Warr et al., 2024; Yan, Sha, et al., 2024).

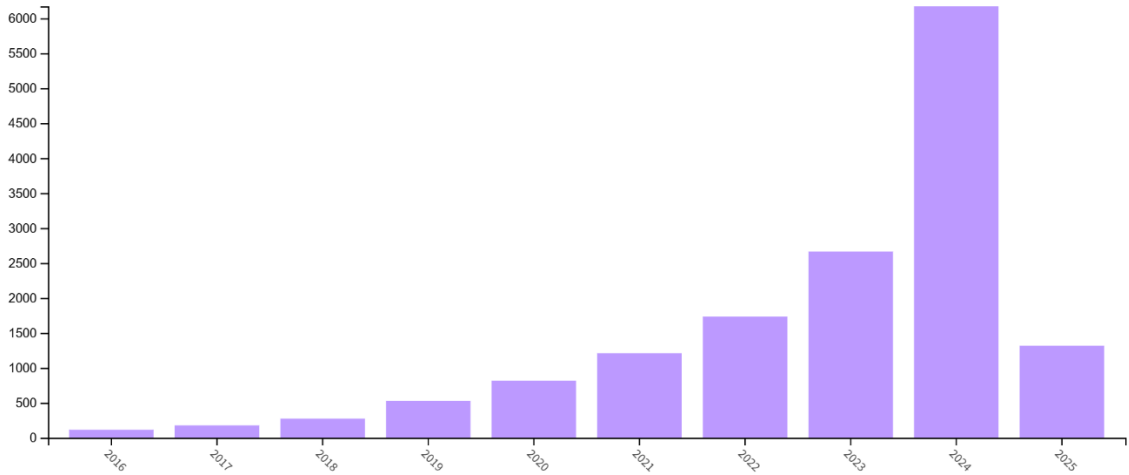

**Fig. 2** Search results from Web of Science using the query related to AI for education.

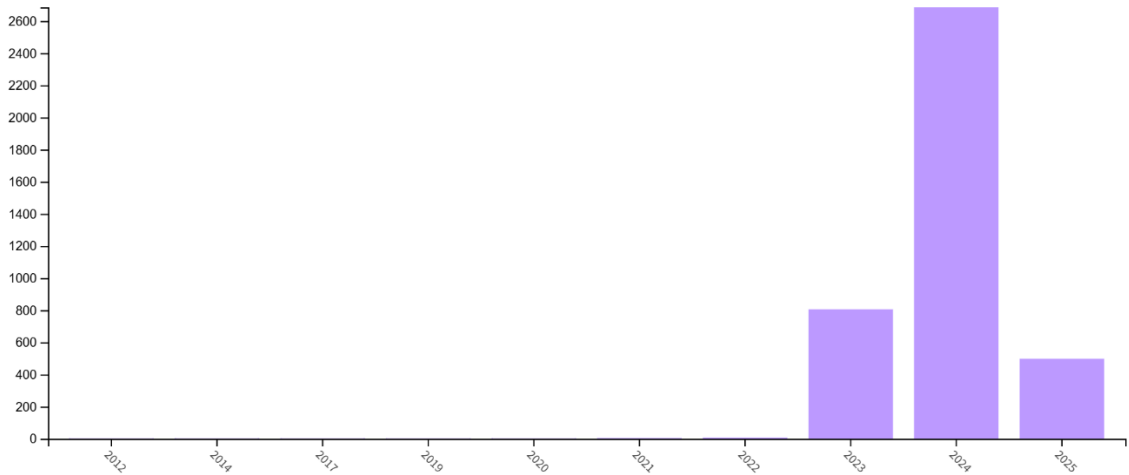

**Fig. 3** Search results from Web of Science using the query related to generative AI (mainly LLMs) in education.

This growing emphasis on LLMs in education raises critical concerns about biases, interpretability, and trust, making it essential to clearly define what kind of AI is being used, for what purpose, and with what ethical safeguards (European Union, 2024; Holmes, 2020; Holmes et al., 2022; Vincent-Lancrin & Van der Vlies, 2020). For instance, one AI application (e.g., LLM-based) might focus on improving learner engagement and motivation, while another might aim to provide personalized learning and teaching. Obviously, these two use cases differ significantly: in the latter, a bias-free, interpretable, and trustworthy AI system is essential to ensure fair and effective personalization. Recently, researchers have fine-tuned LLMs to reduce hallucinations and mitigate biases, improving their reliability (Latif & Zhai, 2024; Sha et al., 2022; Taiye et al., 2024). However, despite these advancements, biases remain inherent in their design (J. Lee et al., 2024; Resnik, 2024). For example, in a mathematics education context, an LLM might learn that algebra is a fundamental branch of math (a factual definition), that students often struggle with word problems (a contingent but neutral fact), and that boys are more likely to excel in advanced math courses (a contingent but problematic bias). If an AI-powered tutor unconsciously reinforces this bias, it may discourage girls from engaging with advanced math by generating responses that subtly reinforce stereotypes—for instance, by suggesting easier problems to female students or offering less encouragement for tackling challenging concepts. Since LLMs cannot inherently distinguish between these categories, these biases remain deeply embedded in their representations, perpetuating stereotypes in ways that are difficult to control or mitigate (A. Chen et al., 2024; J. Lee et al., 2024; Resnik, 2024). Lee et al. (2024) shows that bias in LLM-generated content can emerge at multiple stages of model development, including data curation and fine-tuning, leading to unintended reinforcement of existing disparities. Additionally, the subjective nature of language interpretation in educational settings complicates efforts to define and measure fairness effectively. Moreover, such general AI-based methods still face the broader challenges of deep neural networks, particularly their lack of interpretability (Kasneci et al., 2023). These models do not show the actual decision-making path or provide genuine reasoning behind their outputs. One must be cautious not to mistake LLM-generated explanations for true system-level interpretability, as these models cannot reveal their internal decision-making process. Their immense scale—often spanning tens or hundreds of billions of parameters—makes it infeasible for humans to inspect or comprehend their inner workings, reinforcing their opacity (Singh et al., 2024). With education now recognized as a high-risk area for AI applications, ensuring trustworthiness and interpretability has become an ethical necessity rather than just a technical concern (European Union, 2024). *This underscores the crucial role of custom, domain-specific AI models designed specifically for high-risk areas in education—developed with stakeholder involvement in both design and evaluation. Unlike domain-agnostic, company-driven models optimized purely for metrics without regard for the contextual nature of learning or the complexities of psychological and educational processes, trustworthy and responsible AI for education must be tailored to address existing challenges, ensuring alignment with educational values and equity in learning outcomes.*

### 3.2. Overlooking essential learning processes in learner modelling

Artificial intelligence in education (AIED) has made impressive strides in personalizing instruction, automating feedback, and supporting learners at scale (du Boulay et al., 2023). Central to these advancements is learner modelling (also called student modelling), a fundamental component of AI-driven educational systems that constructs structured representations of learners' cognitive and/or non-cognitive characteristics (Du Boulay et al., 2023). This modelling process relies on analysing educational data—such as learner-system interactions—to infer learners' knowledge and learning states (Abyaa et al., 2019; Conati & Lallé, 2023). However, many current systems model learning in overly simplistic ways that overlook essential cognitive, metacognitive, motivational, and emotional processes (Azevedo & Wiedbusch, 2023). This omission has significant implications for the quality, equity, and effectiveness of AI-based educational interventions (Giannakos et al., 2024; Holstein et al., 2019).

A significant issue in current AI-based learner models is the tendency to treat human learning as a static or linear process. Most systems are designed around discrete outcomes such as task completion, knowledge recall, or quiz performance. While these indicators are useful, they fail to capture the dynamic and recursive nature of learning, particularly as students engage in higher-order thinking or attempt to transfer knowledge across contexts (D'Mello & Graesser, 2023). This narrow conceptualization restricts AI's ability to support deeper learning and long-term educational development. Another core limitation is the insufficient attention paid to metacognitive and self-regulated learning (SRL) processes (Azevedo & Wiedbusch, 2023; Molenaar et al., 2023; Winne, 2018). Decades of research in educational psychology have demonstrated the critical role of skills like planning, monitoring, and evaluating one's own learning (Greene et al., 2024). Despite this, AI systems often do not model these behaviours or do so using rigid, rule-based triggers that fail to reflect individual differences in learners' approaches (Järvelä & Hadwin, 2024). As a result, these systems tend to support immediate task performance rather than the development of strategic thinking or lifelong learning skills.

Furthermore, most AIED systems underrepresent the affective and motivational dimensions of learning. Emotions such as confusion, frustration, or interest can significantly shape how students engage with content, persist through challenges, and construct understanding (D'Mello & Graesser, 2012; Pekrun, 2006). Although some systems attempt to detect affect through facial recognition or physiological data, these efforts often rely on noisy inputs and fail to integrate emotional cues into the learning model in meaningful ways (Calvo & D'Mello, 2010; Graesser, 2020). This leads to systems that remain unresponsive or misaligned with the learner's actual experience. Bayesian knowledge tracing (BKT) and deep knowledge tracing (DKT) are among the most effective approaches for learner modelling (Hooshyar & Druzdzel, 2024; Y. Mao, 2018; Piech et al., 2015). However, despite their effectiveness in tracking subject knowledge, these models predominantly rely on system-question and learner-answer sequences, often overlooking essential learning processes such as self-regulatory behaviours. They operate under the assumption that knowledge accrues independently of motivational, emotional, and metacognitive factors. For instance, BKT (Hooshyar & Druzdzel, 2024; Pelánek, 2017) and DKT (Y. Mao, 2018; Piech et al., 2015) primarily monitor learners' knowledge progression but overlook the broader self-regulatory behaviours that influence learning.

Another issue is the underutilization of multimodal and multichannel data (Azevedo et al., in press; Molenaar et al., 2023). Many AI systems continue to rely solely on log files, clickstreams, or response times, even though learners produce rich streams of data through eye tracking, voice, gesture, and more (Azevedo et al., 2019; Blikstein & Worsley, 2016). These additional modalities can provide valuable insights into how students allocate attention, regulate effort, or experience cognitive load, particularly in complex learning tasks. Ignoring this data limits the granularity and fidelity of learning models and contributes to inaccurate assessments of student understanding or engagement (Azevedo et al., in press)

In addition, current AI models often fail to account for the temporal and contextual nature of learning. Learning is not confined to isolated moments; it unfolds over time and is shaped by prior knowledge, socio-cultural background, and environmental factors (Sawyer, 2022). Yet most systems focus on immediate behaviour without situating it within longer-term trajectories or broader learning ecologies. This short-term perspective impedes the development of tools that can support transfer, long-term retention, and equitable outcomes across diverse learner populations (Saint et al., 2022). A further problem is the limited use of theoretical frameworks from cognitive and educational psychology in the design of AI systems. Many learning models are built through data-driven engineering approaches with minimal reference to theories such as cognitive load theory (Sweller, 2023), self-regulated learning frameworks (Panadero, 2017; Winne, 2018), or constructivist learning theories (Sawyer, 2022). This lack of theoretical grounding undermines the interpretability, generalizability, and pedagogical relevance of AI systems.

Finally, the incomplete modelling of learning processes has ethical and equity implications. When systems prioritize performance metrics over learning strategies or deeper understanding, they may disadvantage students who learn differently or who engage in productive struggle (Kapur, 2024). Moreover, systems that fail to model the full range of learner behaviours and experiences risk reinforcing biases or misclassifying students (R. S. Baker & Hawn, 2022). This calls for more inclusive, transparent, and participatory approaches to the design of AI for education (Holstein & Aleven, 2022).

### 3.3. Lack of stakeholder engagement and domain knowledge in AI development

Unlike expert systems (symbolic AI methods), ML models do not explicitly encode domain knowledge, making it difficult for key stakeholders—such as teachers and learning science researchers—to contribute to their design and ensure alignment with educational frameworks (Ilkou & Koutraki, 2020). Instead, ML models rely on discovering patterns in data rather than engaging in explicit reasoning. While training data is essential for recognizing trends and adapting to specific contexts, it may fail to capture the depth of human expertise and domain-specific nuances (R. S. Baker & Hawn, 2022; Besold et al., 2021; Hooshyar et al., 2024). This lack of direct domain knowledge integration poses several challenges in educational AI applications.

First, ML models struggle to integrate disciplinary knowledge, such as psychological constructs, causal relationships, and educators' insights. This limits their ability to reason explicitly about complex educational processes (Hooshyar et al., 2024; Shakya et al., 2021; Tato & Nkambou, 2022). For example, while SRL principles could enhance learner modelling by capturing students' affective states, metacognitive strategies, and behavioural patterns, leading to more meaningful interventions (Järvelä et al., 2023; Türker & Zingel, 2008), the dynamic and adaptive nature of constructs like motivation and emotional regulation makes them difficult to encode in training data alone. Second, because ML models primarily detect statistical patterns, they often prioritize frequently occurring features—regardless of whether those features are causally important, particularly when dealing with limited or unbalanced data. Recent findings by Hooshyar, Azevedo, et al. (2024) indicate that deep learning models trained only on educational datasets may rely on misleading correlations, overlooking essential causal factors in their predictions. Furthermore, Tato and Nkambou (2022) concluded that injecting symbolic domain knowledge can improve the generalizability of deep neural networks and mitigate learning biases of AI models in imbalanced datasets. Educational datasets may fail to reflect the diversity of real-world learning scenarios (R. S. Baker & Hawn, 2022; Torralba & Efros, 2011), which can inadvertently lead to models that reinforce existing inequalities, misguide interventions, or overlook individual learners' needs (Vincent-Lancrin & Van der Vlies, 2020). Finally, ML models rarely provide opportunities to involve key stakeholders—such as educators and learning scientists—in their design, development, and validation (Fu & Weng, 2024). This lack of engagement prevents AI systems from addressing real-world educational challenges and aligning with pedagogical best practices (Celik et al., 2022). However, integrating domain knowledge into ML models can serve as a bridge to stakeholder involvement (Holstein et al., 2019). Embedding educators' expertise ensures AI systems are developed with a deeper understanding of practical learning environments. This collaborative approach enhances the relevance, transparency, and fairness of AI-driven solutions while empowering practitioners to participate actively in system design (Sperling et al., 2024).

For AI to meaningfully support education, explicit methods must be developed to integrate domain knowledge alongside training data (Dash et al., 2022; Garcez & Lamb, 2023; Hooshyar & Yang, 2021). Such integration would help ML models better represent adaptive learning processes, mitigate biases, and improve decision-making. By directly involving stakeholders in AI development, educational systems can foster more equitable, interpretable, and effective learning experiences. One promising approach is the use of Bayesian networks, which offer a hybrid framework that combines expert knowledge with data-driven learning (Pearl, 1988). This makes them particularly effective for latent variable modelling, a crucial aspect of tracking students' evolving cognitive and self-regulatory processes (Hooshyar & Druzdzel, 2024). However, despite their advantages, Bayesian networks can be computationally expensive and/or require carefully defined structures from domain experts, which can be a limiting factor in large-scale applications (Daly et al., 2011).

According to a recent report by Stevens et al. (2020), one of the most critical challenges in developing AI systems is the integration of prior knowledge: "*ML and AI are generally domain-agnostic...Of-the-shelf [ML and AI] practice treats [each of these] datasets in the same way and ignores background knowledge that extends far beyond the raw data...Improving our ability to systematically incorporate diverse forms of background knowledge can impact every aspect of AI.*" Some examples of explicit educational knowledge that could be embedded in AI-driven learning models to support meaningful and effective learner modelling include: i) causal relationships between variables in the training dataset and their connections to established educational concepts—for instance, in modelling metacognitive skills, behaviours triggered by monitoring, judgments of what is learned and what is not, and self-reported measures for exercising metacognitive control are relevant; ii) theoretical frameworks and models that explain relationships between different learning constructs, such as how learners' performance depends not only on subject-specific processes but also on SRL skills, including cognition, metacognition, emotion, and motivation; iii) the importance of specific variables in explaining the phenomena being modelled—for example, the frequency of self-assessment and self-reported metacognition ratings may be more indicative of metacognitive skills than information structuring; iv) the interaction of variables and their impact on learning outcomes, such as the interplay between cognition, metacognition, and emotional state in task performance.

### 3.4. Applying non-sequential machine learning models to temporal data

Sequential data consists of ordered information where the sequence is integral to its meaning, while temporal data is a type of sequential data that includes timestamps, highlighting both order and timing. In educational contexts, sequential data might include a student's sequence of learning activities (e.g., watching a video, taking a quiz, and completing an assignment), whereas temporal data would track the timing of these events, adding a layer of contextual information such as when each activity occurred (R. Baker, 2010; Nath et al., 2024; Peña-Ayala, 2014; Romero et al., 2010; Saint et al., 2022). In contrast, static data consists of independent observations that do not have any inherent order, such as demographic information (age, gender, school level) or test scores that do not relate to each other sequentially. The dependencies in sequential or temporal data capture richer relationships, such as how early learning behaviours influence later outcomes, which are often lost in static datasets (Knight et al., 2017; Reimann, 2009). Traditional ML models encounter significant challenges when dealing with sequential data (Dietterich, 2002), including capturing and leveraging sequential relationships (e.g., long-distance interactions), representing and incorporating complex loss functions, ensuring the learning algorithms are computationally efficient and fast, and feature selection. A key limitation, inherent to their design, is their inability to effectively capture dependencies between ordered elements, making it difficult to model temporal relationships and sequential patterns.

This challenge has led to the development of methods like the sliding window approach (Dietterich, 2002), which transforms sequential data into fixed-size windows to make it compatible with traditional models (Bañeres et al., 2020; Hu et al., 2014; Lavelle-Hill et al., 2024). However, even within the window, the traditional ML models are still ignoring the temporal dynamics and interdependencies critical for understanding sequential patterns (Knight et al., 2017; Nazeri, 2024). Moreover, these models treat each window independently, failing to capture long-term dependencies or sequential correlations outside the fixed range of the window. This lack of holistic context can lead to inaccurate predictions, particularly in complex datasets like those generated in digital learning platforms. For instance, if a student shows declining quiz scores across weeks, a sliding window approach may fail to recognize this trend unless all relevant data fits neatly within a window. Even if all relevant data fits neatly within a window, traditional machine learning methods still fail to account for the temporal order of events and the progression of scores over time, treating the data as a collection of variables rather than a sequential trend. Similarly, recurrent sliding windows (Bakiri & Dietterich, 1999), which incorporate previously predicted values as inputs to subsequent predictions, compound errors over time, further hindering efficiency and accuracy in modelling long-term dependencies. Other traditional approaches, such as hidden Markov models (Eddy, 1996) or conditional random fields (Sutton & McCallum, 2012), provide better modelling of sequence structures but are computationally expensive and assume simplistic relationships between states (e.g., Markov property), limiting their applicability for complex educational scenarios. Furthermore, these models struggle to incorporate complex loss functions (Dietterich, 2002). For example, when predicting whether a student is at risk of failing, the loss function might penalize the model more for missing early signs of failure, as early intervention is more critical. Traditional models, however, tend to treat all prediction errors equally, regardless of when they occur, which makes it difficult for them to handle loss functions that prioritize early detection over later-stage predictions.

Neural networks, particularly those designed to handle sequential data like Recurrent Neural Networks (RNNs) (Rumelhart et al., 1986) and their variants like Long Short-Term Memory networks (LSTMs) (Hochreiter, 1997), offer significant advantages in addressing these limitations. Unlike coupling fixed-size windows with traditional ML methods, they are designed to learn temporal dynamics, interdependencies, and long-term dependencies within sequential data. RNNs can inherently predict outcomes at any point in a sequence while maintaining the context of the entire sequence, making them versatile for dynamic tasks. LSTMs, additionally, use gating mechanisms to control the flow of information, enabling them to retain critical information from earlier time steps and discard irrelevant details. This makes them particularly effective for tasks such as predicting a student's performance based on a sequence of prior activities spread across different timeframes (Waheed et al., 2023). Additionally, these advanced neural architectures support flexible loss functions, allowing the incorporation of domain-specific priorities, such as penalizing early prediction errors more heavily to support timely interventions. For instance, in predicting student dropout risk, classic sequential models might focus on recent activities, such as low quiz scores, missed assignments, or reduced login frequency, without considering long-term engagement trends. This can result in missed early indicators of disengagement, such as sporadic participation or struggles with foundational concepts at the start of the course. In contrast, LSTMs can track both immediate behaviours and long-term patterns within the same sequence, recognizing trends like initial struggles followed by gradual improvement, which traditional methods would fail to capture. This ability to model long-term dependencies, incorporate flexible loss functions, and efficiently process sequence-level patterns makes deep neural networks indispensable for analysing sequential and temporal data in education. They are also computationally optimized through techniques like parallel processing on GPUs, enabling efficient training and inference.

Finally, feature selection remains a critical challenge in classic sequential/temporal supervised learning, as identifying the most relevant subset of information for making predictions is significantly more complex than in standard supervised learning. The wrapper approach (Kohavi & John, 1997), which generates subsets of features through forward selection or backward elimination and evaluates them via cross-validation or the Akaike information criterion (Sakamoto et al., 1986), is computationally infeasible in long sequences due to the vast number of possible feature combinations. Penalization-based methods, including ridge regression (Hoerl & Kennard, 1970), neural network weight elimination (Weigend et al., 1990), and L1-norm support vector machines (SVMs) (Hearst et al., 1998), shrink the parameters of less relevant features but may still fail to capture long-distance interactions. Feature relevance measures, such as mutual information, provide a simple way to eliminate low-scoring features but do not account for interactions between them (Chow & Liu, 1968; Quinlan, 2014). More advanced techniques, such as RELIEFF (Kononenko et al., 1997), Markov blankets (Koller & Sahami, 1996), and feature racing (Maron & Moore, 1993), attempt to address feature dependencies but remain susceptible to overfitting in long sequences. Finally, tree-structured Bayesian networks (Chow & Liu, 1968), while efficient for identifying low-influence features, may oversimplify complex relationships in sequential data. For example, when predicting student performance, selecting the most relevant features like quiz scores, time spent on assignments, and forum interactions is crucial, but traditional methods may fail to capture the evolving relationships between early struggles and later performance, leading to inaccurate predictions. In contrast, sequential deep neural networks, RNNs and LSTMs, inherently perform feature selection by learning hierarchical representations of relevant features over time, effectively addressing the limitations of manual feature selection methods. Despite the challenges, there are still many research studies employing traditional ML models to process sequential or temporal data in education (Kovacic, 2010; S. Lee & Chung, 2019; Van Petegem et al., 2023; Zhidkikh et al., 2024).

**3.5. Misusing non-sequential metrics to evaluate sequential or temporal models**

Evaluating the performance of models in ML requires robust quantitative metrics. In classification problems, metrics like accuracy, precision, recall, F1-score, and the area under the curve (AUC) are widely used to measure a model's ability to correctly differentiate between categories. For regression tasks, performance is typically assessed using metrics such as root mean squared error (RMSE) and mean absolute error (MAE), which capture the extent of deviation between predicted outcomes and actual values (Kotu & Deshpande, 2014). However, these traditional performance metrics assume that data points are independent and identically distributed, which is not the case in sequential or temporal models. When assessing models designed for time-dependent data, such as RNNs or attention-based models, an evaluation metric should account for sequential dependencies, temporal consistency, the stability of predictions over time, and the points within the sequence where misclassifications or errors occur, as these errors can have different implications depending on their position. For instance, misclassifying a learner's knowledge state in the early sequences may lead to incorrect long-term predictions and improper adaptation of learning materials, whereas errors later in the sequence might indicate instability or overfitting to recent patterns rather than accurately capturing the overall learning trajectory.

Despite these needs, many researchers still rely on non-sequential metrics such as F1-score, accuracy, and AUC when evaluating sequential models, such as DKT (Hooshyar et al., 2022b; Y. Mao, 2018; Piech et al., 2015). While these metrics provide insights into overall predictive performance, they fail to measure the stability of predictions over time. For instance, in DKT, accuracy and F1-score indicate how well the model predicts a student's response to the next question, but they do not capture whether the model's predictions fluctuate erratically across time steps. This means that a model could achieve high F1-scores while producing unstable predictions that vary unpredictably for similar sequences of student responses (Abdelrahman et al., 2023; Yeung & Yeung, 2018). In educational predictive modelling, including DKT, researchers often supplement quantitative metrics with qualitative approaches, such as heatmap visualizations of prediction waviness or reconstruction plots (Li & Wang, 2023; Pan et al., 2024; Zhao et al., 2023). These visualizations help assess the sequential stability of a model by showing how predictions evolve over time, highlighting inconsistencies that traditional metrics fail to capture. Consequently, relying solely on non-sequential metrics can lead to misleading conclusions about a model's true performance in temporal learning environments.

### 3.6. Using unreliable explainable AI methods to provide explanations for black-box models

ML has shown significant potential in fields like education. However, the complexity and opaque nature of many ML models—commonly described as "black-box"—makes them difficult to interpret. Interpretability in ML refers to making model decisions understandable to humans by uncovering the reasoning behind predictions (Khosravi et al., 2022; Miller, 2019). In line with previous studies (Hooshyar & Yang, 2024b; Molnar, 2020), this work uses *interpretable* and *explainable* interchangeably under the broader term interpretability, distinguishing it from explanation, which pertains specifically to clarifying individual predictions. Furthermore, interpretability is used as an umbrella term encompassing the "*extraction of relevant knowledge from a machine-learning model concerning relationships either contained in data or learned by the model*" (Murdoch et al., 2019). In high-stakes domains like education, where decisions can have profound consequences, interpretability is crucial. It supports informed decision-making by clarifying how features influence predictions, identifying areas for improvement, and guiding targeted interventions, while also fostering trust, uncovering biases, improving fairness, and ensuring accountability (Conati et al., 2018a). To achieve these goals, XAI plays a critical role by addressing the opacity of black-box ML models and providing explanations that enhance transparency in predictions and decision-making.

XAI methods are typically categorized into two main types: intrinsic and post-hoc interpretability (Molnar, 2020). Intrinsically interpretable models—such as decision trees and linear regression—offer transparency by clearly outlining how decisions are made, making them easy to understand. In contrast, post-hoc methods aim to explain the behaviour of more complex models, like deep neural networks, after they have been trained. Although intrinsic models enhance transparency, their simpler structures limit their ability to capture complex, non-linear relationships in data. As a result, they often underperform compared to more advanced models in tasks that demand high predictive accuracy (Hooshyar et al., 2019; Ibrahim & Rusli, 2007; Ilkou & Koutraki, 2020). While post-hoc explanation methods—such as SHAP (Lundberg & Lee, 2017), LIME (Ribeiro et al., 2016), and Grad-CAM (Selvaraju et al., 2017)—are designed to explain black-box models, their own explanation-generation process can be opaque, unstable, or unfaithful, raising concerns about the reliability of the explanations they produce (Hooshyar & Yang, 2024b; Krishna et al., 2022; Lakkaraju & Bastani, 2020; Slack et al., 2020). These methods generate approximations of model behaviour rather than offering direct insight into the decision-making process. For example, LIME builds surrogate models based on local linearity assumptions (Laugel et al., 2018), which can lead to unstable and unreliable explanations, especially in complex datasets (Doumard et al., 2022; Garreau & Luxburg, 2020; Hooshyar & Yang, 2024b). Similarly, while SHAP provides detailed contribution analysis, it struggles with computational efficiency and, like LIME, may generate explanations that omit relevant features or create unrealistic scenarios, raising concerns about reliability (Hooshyar & Yang, 2024b; I. E. Kumar et al., 2020; Slack et al., 2020; Van den Broeck et al., 2022).

This opacity raises concerns about the reliability of explanations. Studies by Hooshyar and Yang (2024b), Lakkaraju and Bastani (2020), and Slack et al. (2020) reveal how post-hoc explanations can mimic black-box predictions without truly reflecting decision-making logic. For example, Hooshyar and Yang (2024b) extracted rules from a knowledge-based neural network and created a corresponding wrapper model. They then applied multiple post-hoc explanation methods, such as SHAP and LIME, to interpret the model. The findings revealed that these post-hoc explanations often fail to accurately reflect the model's decision-making mechanism, frequently misattributing importance to less relevant features. This can foster overconfidence in explanations, misinform stakeholders, and perpetuate errors by emphasizing certain features while omitting or distorting others, leading to misleading conclusions and unrealistic scenarios. To address these challenges, some researchers have explored interpretation methods tailored specifically for neural networks and other complex ML models. These methods leverage internal mechanisms, such as gradients or hidden layer representations, to provide more direct and computationally efficient explanations. For example, Hooshyar, Azevedo, et al. (2024) extracted knowledge from hidden layers to reveal how features and concepts are learned, providing both fine- and coarse-grained explanations of deep neural networks, including latent variables. The model was used to capture key components of learners' computational thinking, with latent variables representing computational thinking skills and concepts—such as debugging, simulation, loops, and conditionals—within an educational game environment. Additionally, Hooshyar and Yang (2024b) examined the inner workings of deep neural networks in the same educational context, comparing knowledge extracted from models to explanations provided by post-hoc methods, uncovering critical gaps in explanation reliability.

Beyond the reliance on opaque post-hoc explanation methods, another challenge is shallow interpretability. For instance, in deep knowledge tracing—many researchers equate the probability of skill mastery generated by these models, often visualized using heatmaps, with interpretability (Duan et al., 2024; Li & Wang, 2023; Pan et al., 2024; Zhao et al., 2023). Alarmingly, only a few studies explicitly attempt to examine the decision-making processes within these models, exploring the internal workings of neural networks to uncover how these predictions are made (Valero-Leal et al., 2023; Xu et al., 2024). This lack of process-based transparency raises critical concerns, as it risks reliance on predictions shaped by spurious correlations or hidden data biases—factors that are often invisible to educators (Hooshyar et al., 2024; Tato & Nkambou, 2022). While skill mastery probabilities offer a convenient summary, they fail to reveal the specific features or interactions driving these predictions, providing educators with little actionable insight into students' learning processes. Addressing this gap requires prioritizing interpretability as a design goal, ensuring that models not only generate predictions but also uncover and communicate the underlying patterns and relationships, thereby fostering trust and enhancing the utility of AI for education (Rudin, 2019).

### 3.7. Overlooking ethical guidelines in addressing data inconsistencies during model training

Data-driven models inherently learn from the patterns in their training data, making them susceptible to biases that can hinder their generalization. *Despite this, many existing studies fail to adhere to established ethical AI guidelines for education* (Porayska-Pomsta et al., 2023), *which emphasize the importance of critically examining datasets before model training*. These guidelines advocate for identifying and mitigating various types of bias—such as representational, measurement, aggregation, and learning biases—to ensure fairness, inclusiveness, and trustworthiness. Such biases can emerge at various points throughout the AI pipeline, including data collection, labelling, model training, evaluation, and deployment (Li et al., 2023, 2025). A common issue is class imbalance, where certain categories have significantly fewer instances, leading to suboptimal performance. For example, in DKT, skills that are rarely mastered or almost always mastered create imbalanced data distributions, making it difficult to accurately predict student performance for those skills.

While technical solutions—such as undersampling, oversampling, or cost-sensitive learning—are frequently used to handle class imbalance, they often address the symptoms rather than the root causes of data inconsistencies. For instance, oversampling methods like SMOTE

(Barros et al., 2019; Chawla et al., 2002) or autoencoders (Baldi, 2012; Hooshyar, 2024) are widely used in educational contexts but come with well-known challenges (Fernández et al., 2018). These include small disjuncts, or tiny clusters of minority class samples, which hinder generalization and increase misclassification rates. Class overlap, where minority and majority class examples mix, can confuse the model and reinforce incorrect patterns. Dataset shift, where training and test distributions differ, reduces model performance on unseen data. Lastly, the curse of dimensionality in high-feature datasets makes distance-based calculations unreliable, often leading to poor-quality synthetic samples and overfitting. Although cost-based learning provides an alternative, it can increase computational complexity and produce less interpretable outcomes. Without grounding these methods in ethical guidelines and principled data practices, such solutions risk becoming ad hoc fixes that fail to address deeper issues of fairness, transparency, and educational relevance. For example, SMOTE may generate synthetic data to balance the class distribution that unintentionally introduces unrealistic student behaviours or characteristics—such as students consistently achieving high scores despite never accessing learning materials, attending classes, or submitting assignments, even though these are prerequisites for academic progress.

To address such challenges, Hooshyar (2024) generated synthetic data using knowledge-enhanced autoencoders, integrating educational knowledge into the loss function to ensure that the generated data aligns with the existing statistical properties of the original data distribution, while preserving meaningful learning patterns. In another attempt, Tato and Nkambou (2022) explored a hybrid approach that integrates expert knowledge using Bayesian networks to address data inconsistency in learner modelling. By leveraging Bayesian inference alongside deep learning, this approach compensates for inconsistencies in training data, improves generalization, and enhances predictive accuracy even in sparse or imbalanced educational datasets. Their method was successfully tested on knowledge tracing tasks in two educational contexts: Logic-Muse, a web-based tutoring system supporting logical reasoning using data from 294 students across 48 exercises; and MorALERT, a serious game assessing socio-moral reasoning with 678 annotated student justifications across five skill levels. Furthermore, Hooshyar and his colleagues (2024) injected symbolic educational knowledge into deep network architectures to compensate for data inconsistency issues and guide the learning process in predicting learner performance in computational thinking tasks. Their findings showed that this approach enhanced model generalizability, outperforming models trained with SMOTE and autoencoder-enriched data, while also reducing reliance on spurious correlations and improving the alignment of learned representations with educationally meaningful patterns. While more research is addressing data inconsistency issues in model training (He et al., 2023; Hooshyar et al., 2024; Tato & Nkambou, 2022), many studies still overlook these challenges and the potential of incorporating domain knowledge in symbolic form to mitigate data inconsistencies and sparsity. As shown by Cui et al. (2024) in Fig. 4, three of the most widely used open-access datasets in deep knowledge tracing research suffer from serious class imbalance—an issue surprisingly overlooked by much of the field. Rather than fundamentally addressing these biases, the focus has often been on developing increasingly complex models. *Alarmingly, when Cui and colleagues re-evaluated three*

*mainstream knowledge tracing models on resampled, balanced test sets, they observed a substantial drop in performance (Fig. 5). This reveals that despite their high accuracy on biased datasets, they heavily rely on answer distribution biases, which limits their generalizability and undermines the goal of accurately modelling students' evolving knowledge states.*

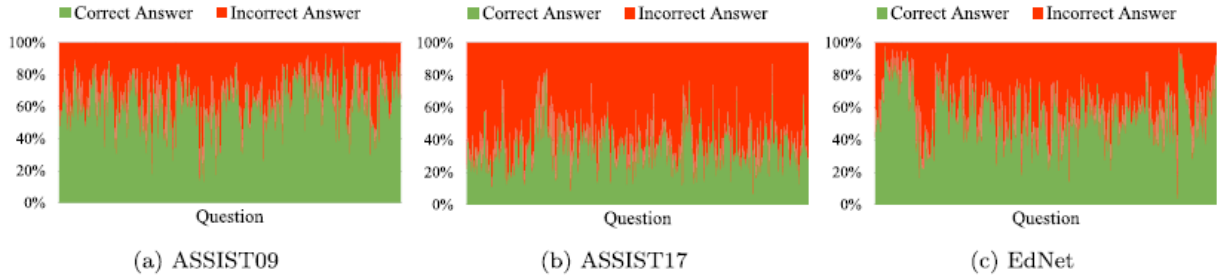

**Fig. 4** Distribution of correct versus incorrect responses per question across three benchmark knowledge tracing datasets, based on a random sample of 500 questions from each dataset, each receiving a minimum of 20 student responses (taken from Cui et al. (2024)).

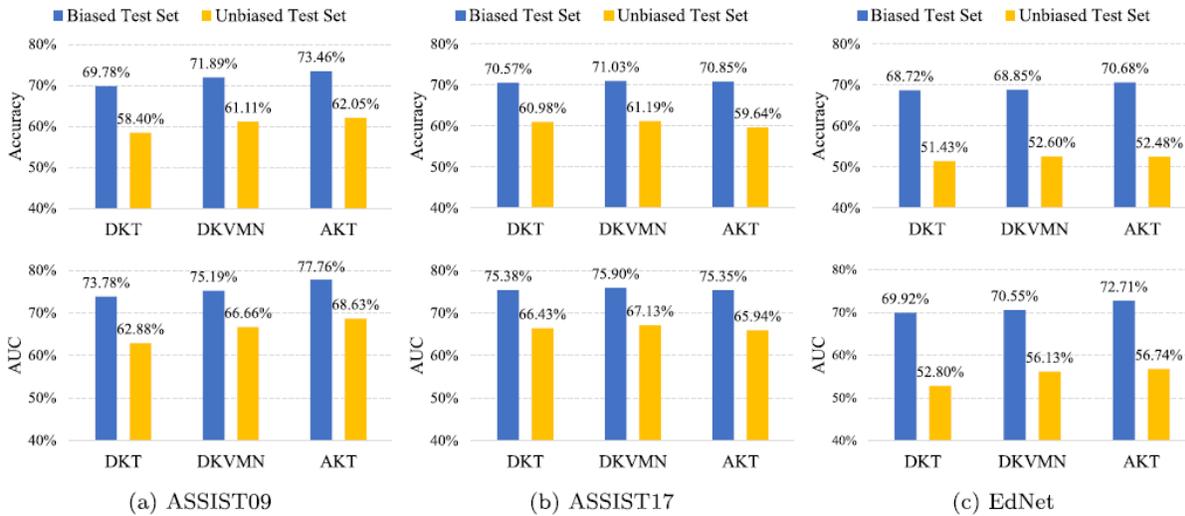

**Fig. 5** Performance variation of DKT, DKVMN, and AKT models on original (biased) versus resampled (unbiased) test sets across multiple datasets (taken from Cui et al. (2024)).

### 3.8. Lack of systematic benchmarking of AI models when identifying patterns

A central objective of AI for education—spanning areas like educational data mining and learning analytics—is to uncover meaningful patterns and key predictors that support data-driven decision-making. These insights span across descriptive, diagnostic, predictive, and prescriptive analytics, with some researchers using identified patterns for diagnosis or intervention strategies, while others focus on predictive modelling to anticipate learning outcomes and trends (R. Baker, 2010; Lang et al., 2022; Romero et al., 2010; Siemens, 2013). While researchers often conduct rigorous model comparisons before deploying models or interpreting predictors (Hooshyar et al., 2020;

Yang et al., 2020), some studies overlook this step, defaulting to mainstream methods without considering alternative AI techniques (Bognár & Fauszt, 2020; Crossley et al., 2020; Divjak et al., 2024; Yun et al., 2025). Without proper model comparison, researchers risk drawing conclusions that may not generalize across different analytical techniques, reducing the robustness of findings in educational research. Addressing this challenge requires systematic benchmarking of AI models before interpreting patterns, ensuring that identified predictors are consistent, reliable, and meaningful within educational contexts.

To this end, Hooshyar et al. (2020) conducted a comparative study on clustering methods for the educational problem of grouping students based on their learning behaviours and performance—using nine student activity datasets extracted from Moodle-based blended learning courses at the University of Tartu—evaluating multiple clustering techniques using various metrics. Their findings reveal that the optimal clustering method depends on dataset characteristics, with DBSCAN and k-medoids performing best for datasets with 10 features, agglomerative and spectral methods excelling with 15 features, and OPTICS ranking highest for larger datasets. Surprisingly, they found that k-means, a widely used method in various research areas, was never ranked the best for any dataset. These results emphasize the importance of rigorous model comparison in educational research and caution against the instinctive use of k-means or other clustering algorithms without thorough evaluation.

In another attempt, Hooshyar et al. (2025) compared symbolic (i.e., decision tree, rule induction, and random forest), sub-symbolic[3] (i.e., logistic regression, Bayesian networks, and deep neural networks), and NSAI (i.e., KBANN) models in terms of generalizability and interpretability. Using datasets on the self-regulated learning behaviours of Estonian primary school students to predict their performance on the 7th-grade national mathematics exam, the researchers found that although both symbolic and sub-symbolic models performed reasonably well on balanced datasets—with some variation in predictive accuracy—they were less effective at detecting low-performing students when applied to imbalanced datasets. The NSAI method, in contrast, proved more generalizable, compensating for data imbalance through its symbolic knowledge component, which improved the representation of underrepresented students. Additionally, in terms of interpretability, symbolic methods primarily highlighted cognitive and motivational factors, while sub-symbolic approaches emphasized cognitive aspects, learned knowledge, and demographic variables but overlooked metacognitive factors. Importantly, different methods within the same AI family identified different variables as the most important predictors in their decision-making. In contrast, NSAI incorporated a broader set of factors—including motivational, cognitive, and metacognitive components, as well as latent representations of knowledge. One main reason for such discrepancies is the fundamental differences in how various AI models learn and process information (Platzer, 2024). Symbolic models operate based on explicit rule-based logic and structured reasoning. Sub-symbolic approaches, like deep neural networks, learn from statistical patterns and correlations, which can cause them to prioritize prominent features—even when those features lack causal significance (Hooshyar et al., 2024). Neural-symbolic models bridge this gap by embedding structured domain knowledge into data-driven learning processes, helping to ensure that the resulting patterns are not only statistically valid but also aligned with theoretical frameworks. This reduces the risk of relying on misleading correlations and enhances the model's ability to capture meaningful relationships (Besold et al., 2021; Garcez & Lamb, 2023; Tato & Nkambou, 2022). Thus, a proper comparison of AI methods should become a standard practice before drawing conclusions, ensuring that the identified predictors and patterns are reliable, theoretically grounded, and not merely artifacts of a specific modelling approach.

---

[3] Sub-symbolic models learn patterns and representations implicitly from data without relying on explicit rules.

### 3.9. Overlooking the potential of local prescriptions

Educational data mining and learning analytics methods widely use ML techniques, whether supervised or unsupervised, to uncover insights that inform stakeholders and support students. These methods include clustering and rule mining, where patterns are often interpretable, as well as regression and classification techniques, which may lack interpretability—particularly in models such as neural networks, SVMs, and some ensemble methods (R. Baker, 2010; Peña-Ayala, 2014; Romero et al., 2010; Siemens, 2013). In such cases, XAI methods are employed to identify learned patterns and enhance transparency. These approaches enable prescriptive analytics, offering general recommendations based on identified trends and relationships. While this significantly aids students and stakeholders, it falls short in providing individualized, case-by-case prescriptions. For instance, assume a trained model reveals that high performers exhibit stronger (meta)cognitive behaviours, such as better information organization, deeper contextual engagement, improved metacognitive awareness, and stronger self-monitoring skills. Additionally, high performers may exhibit greater motivation, measured through persistent behavioural indicators such as the number of task logs per day, the ratio of maximum tasks completed in two hours to total tasks in a day, the number of task submission attempts, and the mean time between consecutive attempts (Toomla et al., 2025).

While models like Bayesian networks can identify the most important predictors of student performance and even reveal dependencies between variables, most ML models lack the capability to provide localized prescriptions. For example, consider a student who exhibits only two dimensions of (meta)cognitive behaviour and one dimension of motivation and falls into a low or medium performance category. A global prescriptive model might tell us which factors matter most in general, but it does not answer a critical local question: *How much should we focus on improving other dimensions of (meta)cognition and motivation to help this specific student reach a high-performance level?* Addressing this gap requires models capable of fine-grained, student-specific recommendations rather than relying solely on global trends.

One prospective solution is employing optimization methods (e.g., genetic algorithms; Holland (1992)) for counterfactual modelling in prescriptive analytics. One example of such an approach is modelling an alternative outcome for high-risk students by determining the minimal set of changes in their input values needed to shift the predictive model's forecast to low risk. To ensure practicality, counterfactual modelling could be constrained to actionable features, excluding immutable characteristics, allowing for more meaningful and implementable recommendations.

For instance, Hooshyar et al. (2022a) proposed a learner modelling approach which automatically analyses students' learning behaviour to predict their risk of failure in courses. Not only does the approach provide explanations for its decisions using XAI methods, fostering trust among educators, practitioners, and learners while supporting global prescriptions, but it also integrates data-driven prescriptive analytics through an evolutionary optimization method to determine the optimal attribute values that maximize the likelihood of low risk of failure. The analytics framework is built on a trained deep learning model applied to a digital literacy course. In this approach, constant values were assigned to the assessment attribute, assuming that every student must take assessments to pass the course. The optimization process prescribes optimal values for other key variables, such as engagement and content access, providing actionable insights to help students adjust their learning behaviours and improve their academic outcomes. In another attempt, Ramaswami et al. (2023) proposed a data-driven prescriptive analytics framework that offers tailored advice to students. Integrated within a learning analytics dashboard, the system suggests specific changes students can implement to improve their learning outcomes. These recommendations are generated using counterfactual analysis, producing individualized, evidence-based advice. The outputs are then translated into clear, human-readable feedback, ensuring that the guidance is both accessible and actionable for students. While some more recent studies have begun adopting this strategy (e.g., Herodotou et al. (2025)), there is still a lack of research exploring optimization methods for delivering personalized, actionable feedback and tailored support prescriptions.

## 4. Hybrid human-AI methods for responsible AI for education

In this section, we describe hybrid human-AI methods and explain how they can address the existing technical and ethical challenges. We begin by introducing the concept of hybrid human-AI intelligence and its relevance in educational contexts. Next, we outline various categorizations of neural-symbolic AI paradigms and discuss how these hybrid approaches tackle key challenges, aligning with the principles of responsible AI for education. Finally, we provide multiple examples to illustrate the practical application of such hybrid methods in educational settings.

### 4.1. Hybrid human-AI intelligence

Hybrid human-AI intelligence has shown great potential for achieving responsible AI for education by combining expert domain knowledge with data-driven methods. This integration ensures that AI systems are not only effective but also transparent and accountable. The human component—often educators, domain experts, or researchers—plays a central role by 1) directly participating in the design and development of AI methods and ensuring that the data-driven learning process is guided by pedagogical principles; and 2) using this pedagogical context to interpret the AI model's behaviour and decision-making, which can, in turn, generate symbolic educational knowledge. This human-in-the-loop process creates a feedback loop in which generated knowledge is used to refine and augment existing domain models, continuously improving both the AI system and the underlying educational frameworks. *Bayesian networks* exemplify this integration by combining expert knowledge with data-driven learning (Pearl, 1988) and offering strong interpretability, particularly for modelling latent student processes (Almond et al., 2015; Hooshyar & Druzdzel, 2024). Specifically, humans can inject domain knowledge in several ways: by designing the structure of the model and allowing data to estimate the parameters; by predefining both the structure and parameters based on expert knowledge; or by learning the structure from data and refining the parameters using human input. These models align well with educational principles and stakeholder expectations, and to a large extent, they are bias-resistant and interpretable—qualities that are crucial for responsible AI in education. However, despite their advantages, Bayesian networks also face limitations—particularly high computational complexity and scalability challenges in large-scale or real-time applications (Daly et al., 2011). In contrast, while deep learning methods address such limitations and typically deliver stronger predictive performance, they often fall short in terms of transparency and the ability to incorporate domain knowledge, which can undermine the trustworthiness of their outputs. *Neural-symbolic AI* (NSAI) has emerged as another promising hybrid approach to overcoming the limitations of traditional deep neural networks (Garcez & Lamb, 2023; Hooshyar & Yang, 2021). Figure 6a provides a general overview of how deep neural networks function, including their use in large language models and educational applications, while Figure 6b illustrates how NSAI can enhance these AI methods for education.

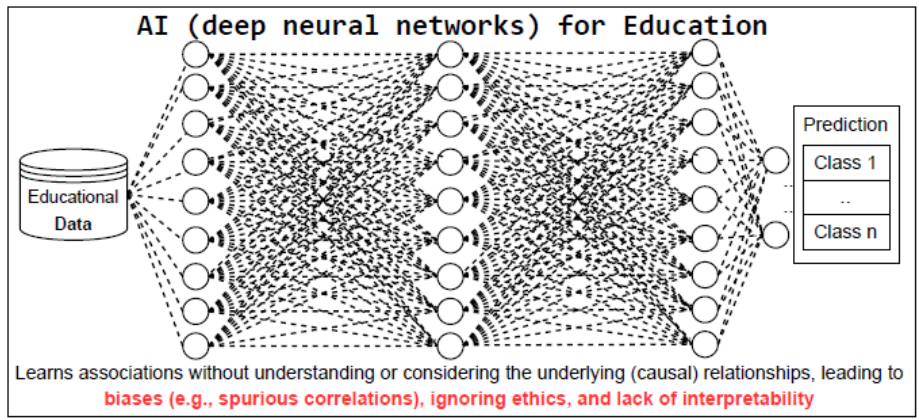

(a)

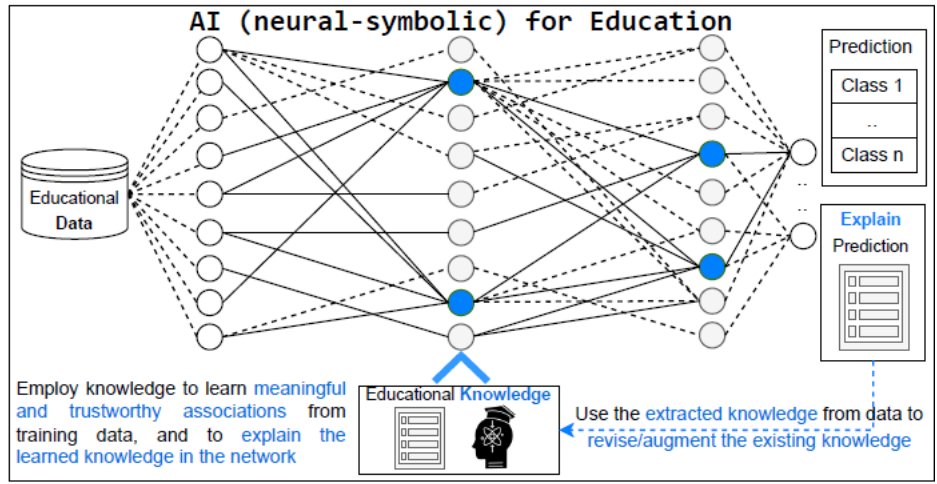

(b)

**Fig. 6** Overview of (a) deep neural networks for education and (b) neural-symbolic AI integration.

### 4.2. Categorization of neural-symbolic AI paradigms

By integrating symbolic reasoning with deep learning's pattern recognition abilities, NSAI bridges the gap between structured knowledge and data-driven learning (Garcez et al., 2022; Sourek et al., 2018). It not only aligns AI-driven insights with educational theories but also enhances generalizability, interpretability, predictive accuracy, and trustworthiness while mitigating bias—key factors for responsible AI for education (Hooshyar et al., 2024; Tato & Nkambou, 2022; Venugopal et al., 2021). Kautz (2022) provided six distinct neural-symbolic AI paradigms, each representing a different way of integrating neural networks and symbolic reasoning. The **Symbolic Neuro-Symbolic** approach, commonly used in natural language processing, processes symbolic input (e.g., words) through neural networks (Pennington et al., 2014). The **Symbolic[Neuro]** architecture embeds a neural pattern recognition module within a symbolic problem solver, as seen in AlphaGo (Silver et al., 2016). In the **Neuro | Symbolic** system, neural networks convert raw data (e.g., images) into symbolic structures, which are then processed by symbolic reasoning engines (J. Mao et al., 2019). The **Neuro: Symbolic → Neuro** model uses symbolic rules for training deep learning models, exemplified by transformer-based symbolic mathematics systems (Lample & Charton, 2019). The **Neuro_{Symbolic}** approach encodes symbolic rules into the structure of neural networks (Serafini et al., 2017), while the **Neuro[Symbolic]** model integrates a symbolic reasoning engine inside a neural network (Kahneman, 2011), aiming to combine symbolic logic with deep learning for enhanced reasoning capabilities. These paradigms reflect the flexibility of NSAI, offering unique opportunities to address specific challenges in educational AI systems.

### 4.3. Addressing key challenges with neural-symbolic AI

Hybrid human-AI methods, such as NSAI, offer a principled and practical response to key challenges of AI in education. By integrating human knowledge and educational concepts into AI models, they ensure decisions are theory-driven rather than solely data-dependent (Kitto et al., 2023)—making it especially well-suited for responsible, trustworthy applications in high-stakes domains like education (European Union, 2024; Fu & Weng, 2024; Pargman et al., 2024; Porayska-Pomsta, 2024). One of the key challenges in AI for education is the limited involvement of teachers, domain experts, and other stakeholders in model development. NSAI addresses this by enabling stakeholders to contribute to model structures through symbolic knowledge injection and to validate and refine outputs via multi-level explainability—from fine- to coarse-grained explanations. This facilitates stakeholder engagement across the entire AI lifecycle—from development to deployment—ensuring models are pedagogically grounded and contextually relevant (Hooshyar et al., 2024, 2025; Tato & Nkambou, 2022). Additionally, NSAI helps counter the neglect of essential learning processes—such as motivation, emotion, and (meta)cognition—by 1) allowing researchers to explicitly model causal and contextual relationships grounded in learning science (Järvelä et al., 2023), and 2) addressing underrepresentation and inconsistencies in data, which often fail to capture these adaptive, temporally evolving processes (R. S. Baker & Hawn, 2022; Hooshyar et al., 2025). Its symbolic component enables the encoding of pedagogical theories alongside behavioural data, resulting in more meaningful and learner-aware models.

Moreover, NSAI supports sequential modelling by embedding contextual and temporal dependencies into its architecture. This addresses limitations of static models and enables temporal patterns to be interpreted in light of educational goals (Shakya et al., 2021; Sourek et al., 2018). Crucially, NSAI also improves explainability. Unlike black-box models that rely on post-hoc interpretation, NSAI offers built-in transparency. Its symbolic layer allows both fine-grained and high-level explanations, giving stakeholders insight into how predictions are made and how decisions align with pedagogical reasoning (Hooshyar et al., 2025; Hooshyar & Yang, 2024b). This directly responds to the challenge of unreliable explanation methods and, indirectly, supports better validation practices by making model behaviour interpretable (Conati et al., 2018b). In terms of ethics and fairness, NSAI supports bias mitigation by enabling the explicit modelling of constraints and fairness principles during training, allowing models to learn meaningful relationships grounded in educational principles rather than spurious correlations (Hooshyar, 2024; Hooshyar et al., 2024; Tato & Nkambou, 2022; Venugopal et al., 2021). This helps ensure that models remain robust even in the face of data inconsistencies or limited training data—an essential step toward ethical, inclusive AI for education. Finally, NSAI allows for both global and local prescriptive analytics. Its structure enables counterfactual reasoning and student-specific recommendations, helping to avoid one-size-fits-all prescriptions and promoting actionable insights tailored to individual learners. Collectively, NSAI provides a robust foundation for transparent, fair, and pedagogically grounded AI for education, fostering human-AI collaboration to advance equity and effectiveness.

Fig. 7 illustrates a simple example of how symbolic knowledge provided by stakeholders could be used to initialize the network structure for more comprehensive and meaningful learner modelling (Hooshyar et al., 2024). The left-side variables or indicators (the small circles) represent our training data, while the oval-shaped variable on the right side represents the label. These digital trace data are all observable and collected during the learning process. We assume that the educational knowledge, in the form of the rules below, is represented using a symbolic knowledge representation language (e.g., propositional logic).

> *Final performance:- Cognition, Metacognition, Emotion, Motivation.*
> *Cognition:- Planning, Search for information, Making inferences, Latent knowledge states.*
> *Latent knowledge states:- Sequences of interactions with tasks and learning materials.*
> *Metacognition:- Goal setting, Information structuring, Judgement of learning.*
> *Information structuring:- Monitoring progress towards goals.*
> *Emotion:- Help seeking, Self-reported emotion rating.*
> *Motivation:- Time watching learning materials, Forum chat.*

For example, the first rule indicates that a student's final performance is influenced by latent variables of *Cognition*, *Metacognition*, *Emotion*, and *Motivation*. Similarly, "Metacognition :- Goal setting, Information structuring, Judgement of learning" means that metacognitive processes are determined by these three observable behaviors. The top and bottom figures (Fig. 7) illustrate the network before and after the training process using training data, respectively. The network is initialized by embedding latent variables using symbolic educational knowledge to capture causal relationships among training variables. In the top figure, green lines indicate fixed connections between final performance and SRL components—cognitive (both domain-general and domain-specific), metacognitive, emotional, and motivational skills. This setup allows the network to integrate SRL skills in its hidden layers, learning latent variables guided by educational knowledge. Fixed, high-weight connections link input variables to latent variables based on symbolic educational rules, such as cognition being influenced by planning, inference-making, and information search. Grey lines represent weakly weighted connections that initially have little effect on the network's output. The bottom figure shows the trained network, where learned adjustments strengthen high-weight connections (solid blue line) between metacognition and time management, as well as cognition and learners' knowledge states. Additionally, negative-weight associations (blue dashed lines) link motivation to automatic logout and emotion to frustration, suggesting that fewer logouts and lower frustration correlate with better affective states.

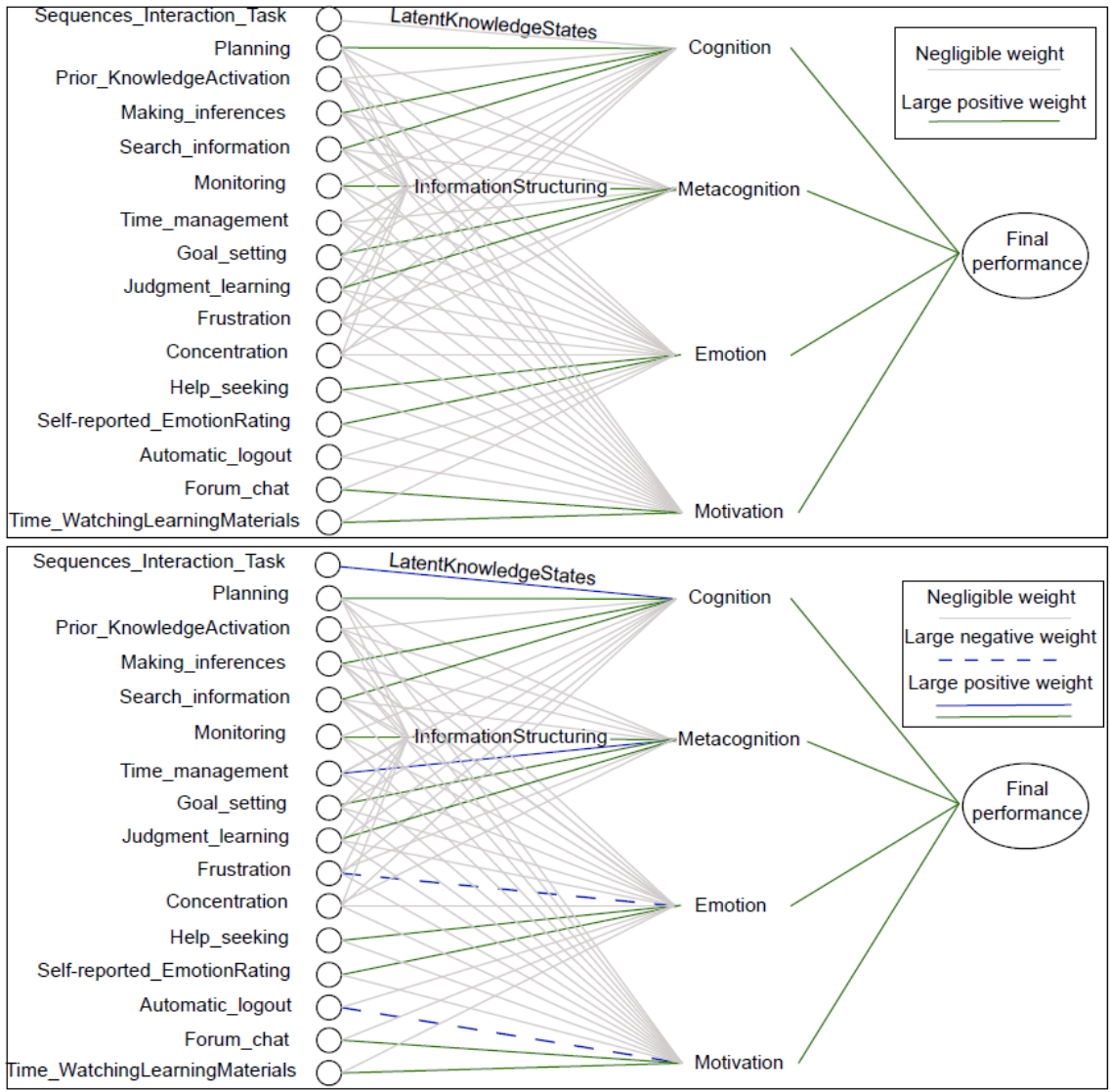

**Fig. 7** Injecting stakeholder knowledge using symbolic knowledge representation (top) and the refined network after training (bottom) (taken from Hooshyar et al. (2024)).

In this realm, Hooshyar and Yang (2021) proposed a general framework for embedding the neural-symbolic computing paradigm into the educational context, which aligns with the *Neuro_{Symbolic}* architecture, serving as a promising foundation for developing interpretable AI for education. The embedding refers to integrating symbolic knowledge (such as explicitly defined educational concepts and rules) directly into neural networks to ensure that both symbolic and neural information can be represented in a compatible way, allowing the network to reason with domain knowledge while learning from data. This enables the development of AI systems that are not only accurate and data-driven but also transparent and aligned with established educational theory, making them more trustworthy and responsible tools for educators and researchers. The proposed framework offers three distinct ways to embed symbolic knowledge into deep neural networks (Dash et al., 2022): modifying network inputs by representing raw features as embeddings that capture both inherent characteristics and domain-specific context, enforcing domain knowledge through penalty terms in the loss function to guide learning, and integrating symbolic knowledge into the network's structure or parameters by defining architectural constraints or introducing priors to model weights (Svatos et al., 2019) (see Fig. 8). In this line, researchers have started to investigate ways to integrate symbolic reasoning with deep learning models. One such effort by Shakya et al. (2021) introduced a hybrid framework that embeds domain-specific educational knowledge—such as the connections between student behaviours and problem-solving strategies—into a deep learning architecture for predicting student strategies. Their approach leveraged Markov Logic Networks (Pedro & Lowd, 2009) to represent structured knowledge and combined this with LSTMs (Hochreiter, 1997) for temporal modelling. To enhance learning efficiency and address overfitting, the model incorporated importance sampling techniques. Experiments using datasets from the KDD EDM challenge highlighted the method's scalability and practical applicability in real-world educational scenarios. In a different vein, Tato and Nkambou (2022) put forward a hybrid approach integrating expert knowledge via Bayesian networks to address data inconsistency in learner modelling. By combining Bayesian inference with deep learning, their method improved generalization and enhanced predictive accuracy, even in sparse or imbalanced datasets. While these efforts represent a significant step towards responsible AI for education, they primarily focus on addressing data inconsistency and biases, with less emphasis on interpretability.

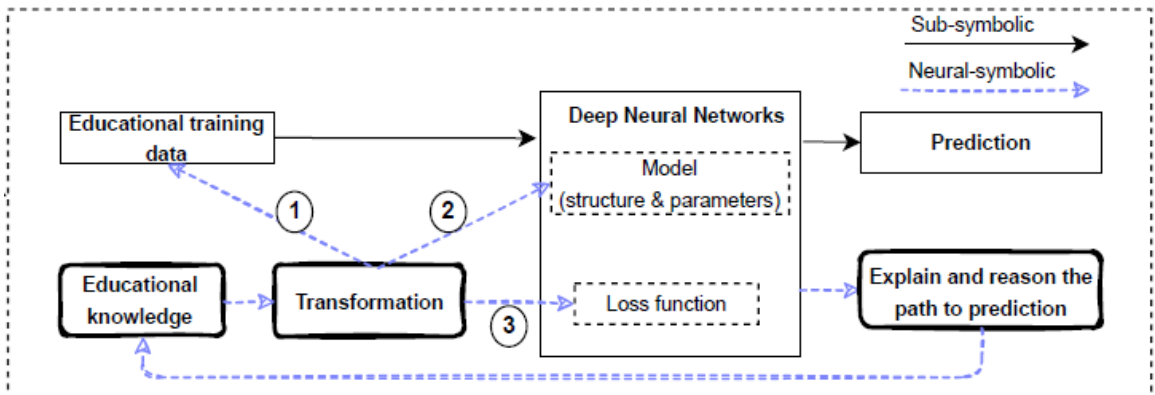

**Fig. 8** The overall process of integrating neural-symbolic AI, where knowledge is both introduced into and extracted from deep neural networks. (1), (2), and (3) illustrate three key approaches for incorporating educational knowledge into the design of deep neural networks.

Building on existing work, Hooshyar and his colleagues have conducted several comprehensive studies (Hooshyar, 2024; Hooshyar et al., 2024; Hooshyar & Yang, 2024b), to explore how neural-symbolic AI can contribute to developing responsible and trustworthy AI for education. Firstly, they embedded educational causal relationships to both guide the learning process and derive interpretable, human-readable rules from the model's outputs (Hooshyar et al., 2024). Their proposed NSAI-based model, which integrates raw student data with symbolic representations of educational principles, demonstrated stronger generalizability and better causal reasoning compared to purely data-driven approaches—even when advanced data augmentation techniques like SMOTE and autoencoders were applied. This work underscores the strength of

NSAI in generating interpretable decision rules, which support structured reasoning and hypothesis validation—an important step toward more responsible practices in educational data mining. In a subsequent effort to evaluate the reliability of explainable AI methods, they analysed four commonly used explainer algorithms in educational contexts. They assessed these methods based on how accurately they reflected the internal logic of trained deep neural networks by extracting and comparing the knowledge embedded within the models (Hooshyar & Yang, 2024b). Their findings highlighted that while the knowledge-enhanced neural network method integrates structured educational knowledge and provides both fine- and coarse-grained interpretability, aligning more closely with educational principles and causal relationships, post-hoc methods such as SHAP and LIME often fall short in accurately capturing a model's decision-making logic. Thirdly, they introduced a novel method that incorporates educational knowledge directly into the loss function of unsupervised deep learning models (Hooshyar, 2024). By embedding symbolic knowledge into autoencoders, the model was guided to penalize behaviours inconsistent with educational guidelines. This integration not only improved predictive performance through synthetic data generation but also preserved alignment with key educational principles. Their findings revealed that this approach improved generalizability compared to models trained on original datasets alone, demonstrating the potential of NSAI for education. Finally, they proposed a responsible educational data mining approach by comparing symbolic, sub-symbolic, and NSAI methods in terms of their interpretability and generalization capabilities within the domain of self-regulated learning (Hooshyar et al., 2025). Their findings revealed that NSAI outperformed other methods in both generalizability and interpretability. By addressing data imbalance and integrating a wider range of SRL factors, NSAI offered a more inclusive and theory-driven educational data mining approach.

Fig. 9a and 9b present the NSAI model's capability to extract both high-level and low-level knowledge related to predicting students' conceptual understanding of mathematics. Fig. 9c provides a global overview of the model's reasoning process (for further details, see Hooshyar et al. (2025)). Unlike conventional ML models—including inherently interpretable ones like decision trees and rule induction, or post-hoc XAI techniques applied to deep neural networks—NSAI enables both theoretical and data-driven interpretability. It achieves this by modelling how observable features influence predictions while also revealing the structure and impact of latent variables. A notable advantage of NSAI is that the extracted explanations are consistent with the educational knowledge embedded during training, allowing researchers to refine and extend theoretical insights. For example, though it was previously understood that unobservable SRL constructs and learned knowledge during testing could predict math conceptual knowledge, Fig. 9a and 9c now provide deeper insights into which specific (latent) variables or constructs play a crucial role. Among them, learned knowledge emerges as the most influential predictor, followed by motivation and cognition, with metacognition and a construct labelled head1[4] having lesser impact. Fig. 9b further demonstrates how observable variables load onto latent constructs.

---

[4] Heads refer to the individual neural network nodes or units within a layer.

Cognitive features show strong associations with the cognition construct, supporting their role in attention regulation, scientific reasoning, and memory-related strategies. Attributes tied to motivation, metacognition, and learned knowledge also contribute meaningfully to their respective latent factors, highlighting their relevance to conceptual knowledge acquisition. Among motivational indicators, growth mindset ranks highest in importance, followed by controlled motivation, calculation behaviour, and help-seeking; meanwhile, autonomous motivation and self-efficacy appear less influential. Moreover, three metacognitive attributes—self-assessment, confidence in learning vocabulary, and the perceived value of association strategies—along with the use of a specific cognitive strategy (association strategy I), contribute significantly to the motivation construct. This pattern suggests a reinforcing interaction between metacognitive and cognitive strategies in shaping students' motivation. Overall, by involving stakeholders in AI design, leveraging domain knowledge to address data inconsistency issues, and extracting knowledge directly from trained networks, NSAI uncovers the influence of various SRL constructs and their underlying components on students' knowledge acquisition during assessments. *This contributes to a deeper, more transparent, and data-informed understanding of learning processes, supporting the advancement of responsible and trustworthy AI for education.*

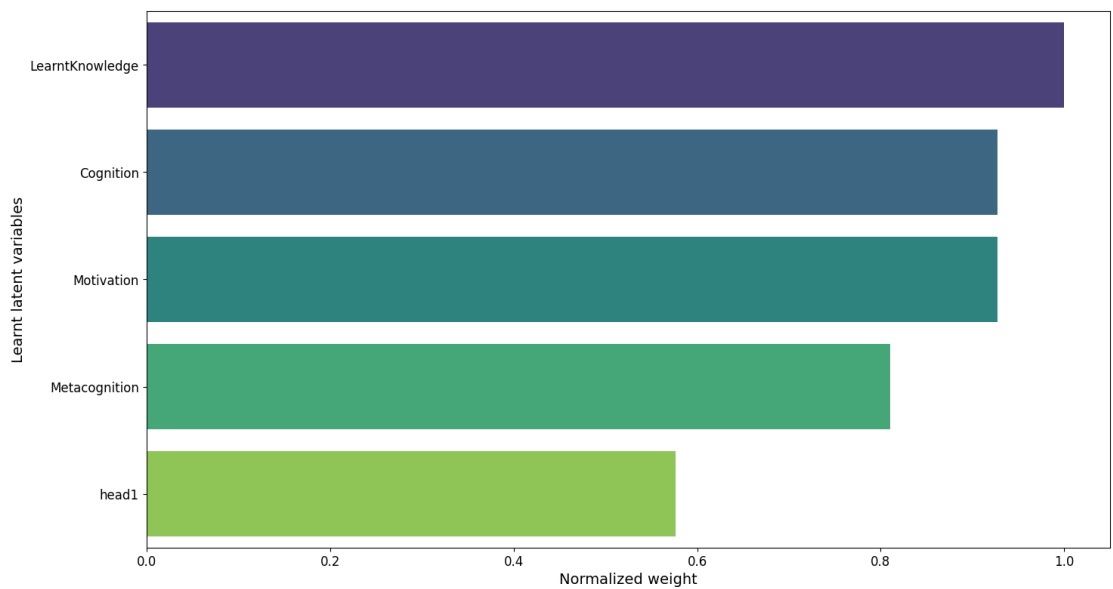

(a)

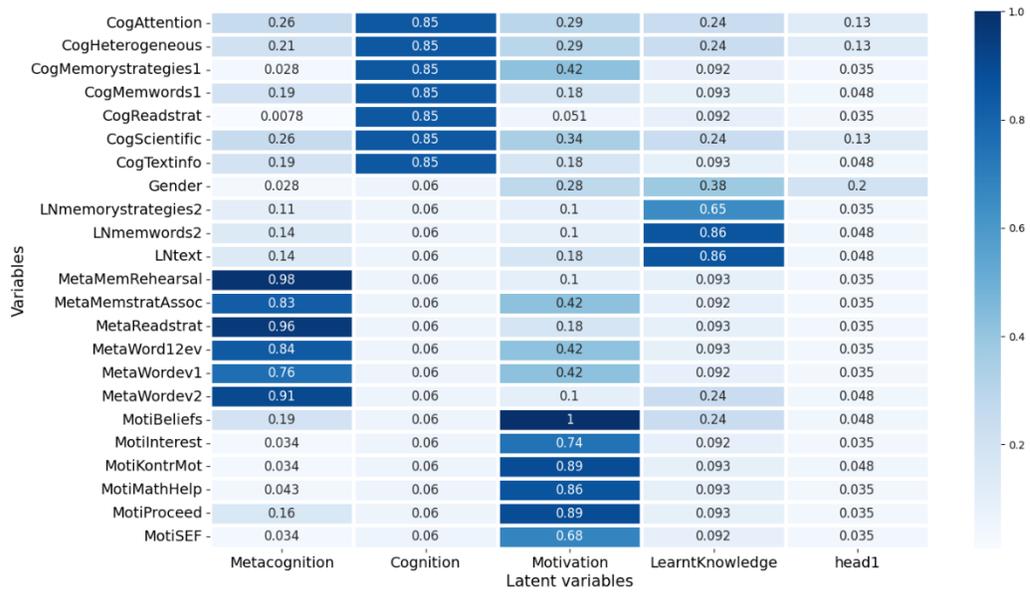

(b)

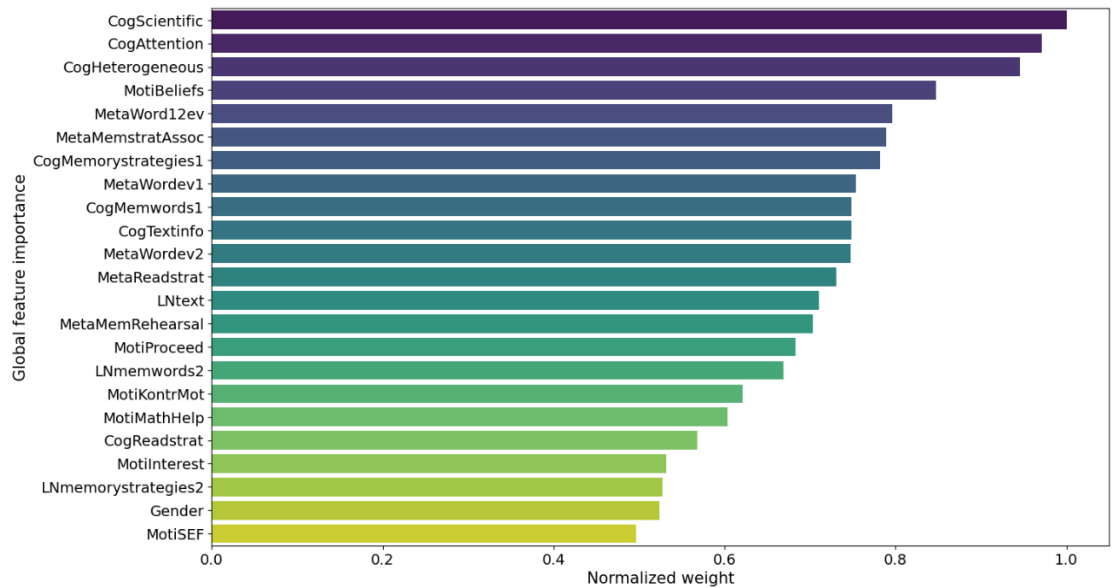

(c)

**Fig. 9** (a) and (b) illustrate high- and low-level knowledge extraction from the NSAI model for predicting math conceptual knowledge, respectively, while (c) presents a global explanation of the learned NSAI model (taken from Hooshyar et al. (2025)).

## 5. Conclusion

Achieving responsible AI for education goes beyond mere adoption—it requires a deliberate and ethical approach to development, ensuring that AI tools align with the needs of students, teachers, and policymakers. Despite advancements in AI-driven educational systems, several critical issues remain unresolved, acting as the *elephants in the room* within the AI in education, learning analytics, educational data mining, learning sciences, and educational psychology communities. These include the oversimplification of AI's role in education, the neglect of meta- and non-cognitive factors in learner modelling, and the limited engagement of educators in AI development. Additionally, issues such as the inappropriate application of machine learning to temporal data, reliance on non-sequential metrics, and the shortcomings in interpretability hinder trust and effectiveness. As AI's role in education expands, the focus must shift from generic, black-box models to purpose-built, context-aware solutions that are transparent, fair, and aligned with educational values.

To address these challenges, we advocate for hybrid AI methods—specifically neural-symbolic AI—as a pathway towards more responsible and trustworthy AI systems. Supported by both theoretical and empirical research, we demonstrate how hybrid AI methods address key challenges in AI for education and serve as the foundation for building responsible, trustworthy AI systems that align with the needs of students, teachers, and policymakers. By embracing hybrid methods such as neural-symbolic AI and systematically addressing the existing challenges, AI for education can move towards more responsible, trustworthy, and effective systems that support meaningful and inclusive learning experiences.

Realizing this potential, however, depends not only on choosing the right technical approach but also on how these hybrid human-AI systems are designed and implemented. In moving towards responsible AI for education, hybrid human-AI systems must be intentionally designed to support, rather than replace, human capabilities. These systems should align with core psychological and educational principles—particularly those that emphasize learner agency, cognitive development, emotional well-being, and self-regulated learning. By integrating human expertise with AI-driven adaptivity, hybrid models can create more personalized, responsive, and ethically grounded learning experiences. This approach ensures that AI acts not as an autonomous decision-maker but as a collaborative partner that augments teachers' instructional practices and scaffolds students' learning processes. From a psychological perspective, responsible hybrid systems must account for the complexity of learning, including the motivational, affective, and metacognitive dimensions often overlooked in purely data-driven designs. By leveraging insights from the learning sciences and human development, AI technologies can better respond to learners' needs, support co-regulation, and foster meaningful engagement. Importantly, these systems must address ethical concerns such as data privacy, transparency, and equity, especially for historically underserved student populations. This commitment to ethical considerations reassures us that AI for education will be used responsibly and for the benefit of all. As we advance, interdisciplinary collaboration among AI researchers, psychologists, computer scientists, educators, technologists, and policymakers are essential to ensure that AI for education remains human-centred, ethically sound, and developmentally appropriate.

Ultimately, the success of hybrid human-AI systems in education will depend on technological innovation and our commitment to psychological integrity and educational values. Responsible AI must be measured by its capacity to empower learners, support teachers, and promote equitable and sustainable educational ecosystems. However, it is important to acknowledge that AI integration in education also poses potential risks and challenges, such as over-reliance on technology and data privacy issues. By prioritizing these goals and addressing these challenges, we can ensure that AI catalyses deeper learning and human flourishing rather than disrupting the educational experience.

## Acknowledgements

This work was supported by the Estonian Research Council grant (PRG2215).

## References

Abdelrahman, G., Wang, Q., & Nunes, B. (2023). Knowledge tracing: A survey. *ACM Computing Surveys*, *55*(11), 1–37.

Abyaa, A., Khalidi Idrissi, M., & Bennani, S. (2019). Learner modelling: Systematic review of the literature from the last 5 years. *Educational Technology Research and Development*, *67*, 1105–1143. https://doi.org/10.1007/s11423-018-09644-1

Aggarwal, C. C. (2018). *Neural networks and deep learning* (Vol. 10, Issue 978). Springer.

Almond, R. G., Mislevy, R. J., Steinberg, L. S., Yan, D., & Williamson, D. M. (2015). *Bayesian networks in educational assessment*. Springer.

Arrieta, A. B., Díaz-Rodríguez, N., Del Ser, J., Bennetot, A., Tabik, S., Barbado, A., García, S., Gil-López, S., Molina, D., & Benjamins, R. (2020). Explainable Artificial Intelligence (XAI): Concepts, taxonomies, opportunities and challenges toward responsible AI. *Information Fusion*, *58*, 82–115.

Azevedo, R., Hooshyar, D., Fan, Y., Wiedbusch, M., & Dever, D. (in press). Multimodal learning analytics for self-regulated learning across diverse learning technologies: Analysis, prediction, generative AI, and Explainable AI. In *Handbook of online learning measures*. European Association for Research on Learning and Instruction (EARLI).

Azevedo, R., Mudrick, N. V., Taub, M., & Bradbury, A. E. (2019). *Self-regulation in computer-assisted learning systems*.

Azevedo, R., & Wiedbusch, M. (2023). Theories of metacognition and pedagogy applied to AIED systems. In *Handbook of artificial intelligence in education* (pp. 45–67). Edward Elgar Publishing.

Baker, R. (2010). Data mining for education. *International Encyclopedia of Education*, *7*(3), 112–118.

Baker, R. S., & Hawn, A. (2022). Algorithmic bias in education. *International Journal of Artificial Intelligence in Education*, 1–41.

Bakiri, G., & Dietterich, T. G. (1999). Achieving high-accuracy text-to-speech with machine learning. *Data Mining in Speech Synthesis*, *10*.

Baldi, P. (2012). *Autoencoders, unsupervised learning, and deep architectures*. 37–49.
Bañeres, D., Rodríguez, M. E., Guerrero-Roldán, A. E., & Karadeniz, A. (2020). An early warning system to detect at-risk students in online higher education. *Applied Sciences*, *10*(13), 4427.
Barros, T. M., Souza Neto, P. A., Silva, I., & Guedes, L. A. (2019). Predictive models for imbalanced data: A school dropout perspective. *Education Sciences*, *9*(4), 275.
Besold, T. R., d'Avila Garcez, A., Bader, S., Bowman, H., Domingos, P., Hitzler, P., Kühnberger, K.-U., Lamb, L. C., Lima, P. M. V., & de Penning, L. (2021). Neural-symbolic learning and reasoning: A survey and interpretation 1. In *Neuro-Symbolic Artificial Intelligence: The State of the Art* (pp. 1–51). IOS press.
Blikstein, P., & Worsley, M. (2016). Multimodal learning analytics and education data mining: Using computational technologies to measure complex learning tasks. *Journal of Learning Analytics*, *3*(2), 220–238.
Bognár, L., & Fauszt, T. (2020). *Different learning predictors and their effects for Moodle Machine Learning models*. 000405–000410.
Breiman, L., Friedman, J., Stone, C., & Olshen, R. (1984). Classification and Regression Trees: Taylor & Francis. *Taylor Francis*.
Calvo, R. A., & D'Mello, S. (2010). Affect detection: An interdisciplinary review of models, methods, and their applications. *IEEE Transactions on Affective Computing*, *1*(1), 18–37.
Celik, I., Dindar, M., Muukkonen, H., & Järvelä, S. (2022). The promises and challenges of artificial intelligence for teachers: A systematic review of research. *TechTrends*, *66*(4), 616–630.
Chawla, N. V., Bowyer, K. W., Hall, L. O., & Kegelmeyer, W. P. (2002). SMOTE: synthetic minority over-sampling technique. *Journal of Artificial Intelligence Research*, *16*, 321–357.
Chen, A., Malladi, S., Zhang, L., Chen, X., Zhang, Q. R., Ranganath, R., & Cho, K. (2024). Preference learning algorithms do not learn preference rankings. *Advances in Neural Information Processing Systems*, *37*, 101928–101968.
Chen, L., Chen, P., & Lin, Z. (2020). Artificial intelligence in education: A review. *Ieee Access*, *8*, 75264–75278.
Chow, C., & Liu, C. (1968). Approximating discrete probability distributions with dependence trees. *IEEE Transactions on Information Theory*, *14*(3), 462–467.
Clancey, W. J. (1979). Tutoring rules for guiding a case method dialogue. *International Journal of Man-Machine Studies*, *11*(1), 25–49.
Conati, C., & Lallé, S. (2023). 8. Student modeling in open-ended learning environments. *Handbook of Artificial Intelligence in Education*, 170–183.
Conati, C., Porayska-Pomsta, K., & Mavrikis, M. (2018a). AI in Education needs interpretable machine learning: Lessons from Open Learner Modelling. *arXiv Preprint arXiv:1807.00154*.
Conati, C., Porayska-Pomsta, K., & Mavrikis, M. (2018b). AI in Education needs interpretable machine learning: Lessons from Open Learner Modelling. *arXiv Preprint arXiv:1807.00154*. https://doi.org/10.48550/arXiv.1807.00154
Crossley, S. A., Karumbaiah, S., Ocumpaugh, J., Labrum, M. J., & Baker, R. S. (2020). Predicting math identity through language and click-stream patterns in a blended learning mathematics program for elementary students. *Journal of Learning Analytics*, *7*(1), 19–37.

Cui, C., Ma, H., Dong, X., Zhang, C., Zhang, C., Yao, Y., Chen, M., & Ma, Y. (2024). Model-agnostic counterfactual reasoning for identifying and mitigating answer bias in knowledge tracing. *Neural Networks*, *178*, 106495.

Daly, R., Shen, Q., & Aitken, S. (2011). Learning Bayesian networks: Approaches and issues. *The Knowledge Engineering Review*, *26*(2), 99–157.

Dash, T., Chitlangia, S., Ahuja, A., & Srinivasan, A. (2022). A review of some techniques for inclusion of domain-knowledge into deep neural networks. *Scientific Reports*, *12*(1), 1040. https://doi.org/10.1038/s41598-021-04590-0

Dhar, V. (2024). The paradigm shifts in artificial intelligence. *Communications of the ACM*, *67*(11), 50–59.

Dietterich, T. G. (2002). *Machine learning for sequential data: A review*. 15–30.

Divjak, B., Svetec, B., & Horvat, D. (2024). How can valid and reliable automatic formative assessment predict the acquisition of learning outcomes? *Journal of Computer Assisted Learning*, *40*(6), 2616–2632.

D'Mello, S., & Graesser, A. (2012). Dynamics of affective states during complex learning. *Learning and Instruction*, *22*(2), 145–157.

D'Mello, S. K., & Graesser, A. (2023). Intelligent tutoring systems: How computers achieve learning gains that rival human tutors. In *Handbook of educational psychology* (pp. 603–629). Routledge.

Doumard, E., Aligon, J., Escriva, E., Excoffier, J.-B., Monsarrat, P., & Soulé-Dupuy, C. (2022). *A comparative study of additive local explanation methods based on feature influences*. *3130*(paper 4), 31–40.

Du Boulay, B., Mitrovic, A., & Yacef, K. (2023). *Handbook of artificial intelligence in education*. Edward Elgar Publishing.

Duan, Z., Dong, X., Gu, H., Wu, X., Li, Z., & Zhou, D. (2024). Towards more accurate and interpretable model: Fusing multiple knowledge relations into deep knowledge tracing. *Expert Systems with Applications*, *243*, 122573.

Durán, J. M., & Jongsma, K. R. (2021). Who is afraid of black box algorithms? On the epistemological and ethical basis of trust in medical AI. *Journal of Medical Ethics*, *47*(5), 329–335.

Eddy, S. R. (1996). Hidden markov models. *Current Opinion in Structural Biology*, *6*(3), 361–365.

Eitel-Porter, R. (2020). *Beyond the promise: Implementing ethical AI*.

European Union. (2024). *Artificial Intelligence Act*. https://artificialintelligenceact.eu/

Fernández, A., Garcia, S., Herrera, F., & Chawla, N. V. (2018). SMOTE for learning from imbalanced data: Progress and challenges, marking the 15-year anniversary. *Journal of Artificial Intelligence Research*, *61*, 863–905.

Forero-Corba, W., & Bennasar, F. N. (2024). Techniques and applications of Machine Learning and Artificial Intelligence in education: A systematic review. *RIED-Revista Iberoamericana de Educación a Distancia*, *27*(1).

Fu, Y., & Weng, Z. (2024). Navigating the ethical terrain of AI in education: A systematic review on framing responsible human-centered AI practices. *Computers and Education: Artificial Intelligence*, 100306.

Garcez, A. d'Avila, Bader, S., Bowman, H., Lamb, L. C., de Penning, L., Illuminoo, B., Poon, H., & Zaverucha, C. G. (2022). Neural-symbolic learning and reasoning: A survey and interpretation. *Neuro-Symbolic Artificial Intelligence: The State of the Art*, *342*(1), 327. https://doi.org/10.48550/arXiv.1711.03902
Garcez, A. d'Avila, & Lamb, L. C. (2023). Neurosymbolic AI: The 3 rd wave. *Artificial Intelligence Review*, 1–20.
Garreau, D., & Luxburg, U. (2020). *Explaining the explainer: A first theoretical analysis of LIME*. 1287–1296.
Gervet, T., Koedinger, K., Schneider, J., & Mitchell, T. (2020). When is deep learning the best approach to knowledge tracing? *Journal of Educational Data Mining*, *12*(3), 31–54.
Giannakos, M., Azevedo, R., Brusilovsky, P., Cukurova, M., Dimitriadis, Y., Hernandez-Leo, D., Järvelä, S., Mavrikis, M., & Rienties, B. (2024). The promise and challenges of generative AI in education. *Behaviour & Information Technology*, 1–27.
Goellner, S., Tropmann-Frick, M., & Brumen, B. (2024). Responsible Artificial Intelligence: A Structured Literature Review. *arXiv Preprint arXiv:2403.06910*.
Goodfellow, I., Pouget-Abadie, J., Mirza, M., Xu, B., Warde-Farley, D., Ozair, S., Courville, A., & Bengio, Y. (2020). Generative adversarial networks. *Communications of the ACM*, *63*(11), 139–144.
Goodman, B., & Flaxman, S. (2017). European Union regulations on algorithmic decision-making and a "right to explanation." *AI Magazine*, *38*(3), 50–57.
Graesser, A. C. (2020). Emotions are the experiential glue of learning environments in the 21st century. *Learning and Instruction*, *70*, 101212.
Greene, J. A., Bernacki, M. L., & Hadwin, A. F. (2024). Self-regulation. *Handbook of Educational Psychology*, 314–334.
Hatzilygeroudis, I., & Prentzas, J. (2004). Neuro-symbolic approaches for knowledge representation in expert systems. *International Journal of Hybrid Intelligent Systems*, *1*(3–4), 111–126.
Havrilla, A., Du, Y., Raparthy, S. C., Nalmpantis, C., Dwivedi-Yu, J., Zhuravinskyi, M., Hambro, E., Sukhbaatar, S., & Raileanu, R. (2024). Teaching large language models to reason with reinforcement learning. *arXiv Preprint arXiv:2403.04642*.
He, L., Li, X., Wang, P., Tang, J., & Wang, T. (2023). Integrating fine-grained attention into multi-task learning for knowledge tracing. *World Wide Web*, *26*(5), 3347–3372.
Hearst, M. A., Dumais, S. T., Osuna, E., Platt, J., & Scholkopf, B. (1998). Support vector machines. *IEEE Intelligent Systems and Their Applications*, *13*(4), 18–28.
Heaton, D., Nichele, E., Clos, J., & Fischer, J. E. (2023). "The algorithm will screw you": Blame, social actors and the 2020 A Level results algorithm on Twitter. *Plos One*, *18*(7), e0288662.
Herodotou, C., Carr, J., Shrestha, S., Comfort, C., Bayer, V., Maguire, C., Lee, J., Mulholland, P., & Fernandez, M. (2025). *Prescriptive analytics motivating distance learning students to take remedial action: A case study of a student-facing dashboard*. 306–316.
Hinton, G. E. (1992). How neural networks learn from experience. *Scientific American*, *267*(3), 144–151.
Hochreiter, S. (1997). *Long Short-Term Memory| Neural Computation| MIT Press*.
Hoerl, A. E., & Kennard, R. W. (1970). Ridge regression: Biased estimation for nonorthogonal problems. *Technometrics*, *12*(1), 55–67.
Hoffmann, J., Borgeaud, S., Mensch, A., Buchatskaya, E., Cai, T., Rutherford, E., Casas, D. de L., Hendricks, L. A., Welbl, J., & Clark, A. (2022). Training compute-optimal large language models. *arXiv Preprint arXiv:2203.15556*.

Holland, J. H. (1992). Genetic algorithms. *Scientific American*, *267*(1), 66–73.
Holmes, W. (2020). Artificial intelligence in education. In *Encyclopedia of education and information technologies* (pp. 88–103). Springer. https://doi.org/10.1007/978-3-030-10576-1_107
Holmes, W., Persson, J., Chounta, I.-A., Wasson, B., & Dimitrova, V. (2022). *Artificial intelligence and education: A critical view through the lens of human rights, democracy and the rule of law*. Council of Europe. https://discovery.ucl.ac.uk/id/eprint/10158376/
Holstein, K., & Aleven, V. (2022). Designing for human–AI complementarity in K-12 education. *AI Magazine*, *43*(2), 239–248.
Holstein, K., McLaren, B. M., & Aleven, V. (2019). Co-designing a real-time classroom orchestration tool to support teacher-AI complementarity. *Grantee Submission*.
Hooshyar, D. (2024). Temporal learner modelling through integration of neural and symbolic architectures. *Education and Information Technologies*, *29*(1), 1119–1146. https://doi.org/10.1007/s10639-023-12334-y
Hooshyar, D., Azevedo, R., & Yang, Y. (2024). Augmenting Deep Neural Networks with Symbolic Educational Knowledge: Towards Trustworthy and Interpretable AI for Education. *Machine Learning and Knowledge Extraction*, *6*(1), 593–618. https://doi.org/10.3390/make6010028
Hooshyar, D., & Druzdzel, M. J. (2024). Memory-Based Dynamic Bayesian Networks for Learner Modeling: Towards Early Prediction of Learners' Performance in Computational Thinking. *Education Sciences*, *14*(8), 917.
Hooshyar, D., Huang, Y.-M., & Yang, Y. (2022a). A Three-Layered Student Learning Model for Prediction of Failure Risk in Online Learning. *Human-Centric Computing and Information Sciences*, *12*. https://doi.org/10.22967/HCIS.2022.12.028
Hooshyar, D., Huang, Y.-M., & Yang, Y. (2022b). GameDKT: Deep knowledge tracing in educational games. *Expert Systems with Applications*, *196*, 116670. https://doi.org/10.1016/j.eswa.2022.116670
Hooshyar, D., Kikas, E., Yang, Y., Šír, G., Hämäläinen, R., Kärkkäinen, T., & Azevedo, R. (2025). Towards Responsible and Trustworthy Educational Data Mining: Comparing Symbolic, Sub-Symbolic, and Neural-Symbolic AI Methods. *arXiv Preprint arXiv.2504.00615*.
Hooshyar, D., Pedaste, M., & Yang, Y. (2019). Mining educational data to predict students' performance through procrastination behavior. *Entropy*, *22*(1), 12. https://doi.org/10.3390/e22010012
Hooshyar, D., & Yang, Y. (2021). Neural-symbolic computing: A step toward interpretable AI in education. *Bulletin of the Technical Committee on Learning Technology (ISSN: 2306-0212)*, *21*(4), 2–6. https://tc.computer.org/tclt/10-1109-2021-0401011/
Hooshyar, D., & Yang, Y. (2024a). ImageLM: Interpretable image-based learner modelling for classifying learners' computational thinking. *Expert Systems with Applications*, *238*, 122283. https://doi.org/10.1016/j.eswa.2023.122283
Hooshyar, D., & Yang, Y. (2024b). Problems with SHAP and LIME in interpretable AI for education: A comparative study of post-hoc explanations and neural-symbolic rule extraction. *IEEE Access*.
Hooshyar, D., Yang, Y., Pedaste, M., & Huang, Y.-M. (2020). Clustering algorithms in an educational context: An automatic comparative approach. *IEEE Access*, *8*, 146994–147014.

Hu, Y.-H., Lo, C.-L., & Shih, S.-P. (2014). Developing early warning systems to predict students' online learning performance. *Computers in Human Behavior*, *36*, 469–478.
Ibrahim, Z., & Rusli, D. (2007). *Predicting students' academic performance: Comparing artificial neural network, decision tree and linear regression*. 21st Annual SAS Malaysia Forum, 5th September.
Ilkou, E., & Koutraki, M. (2020). *Symbolic vs sub-symbolic ai methods: Friends or enemies? 2699*.
İpek, Z. H., Gözüm, A. I. C., Papadakis, S., & Kallogiannakis, M. (2023). Educational Applications of the ChatGPT AI System: A Systematic Review Research. *Educational Process: International Journal*, *12*(3), 26–55.
Jakesch, M., Buçinca, Z., Amershi, S., & Olteanu, A. (2022). *How different groups prioritize ethical values for responsible AI*. 310–323.
Järvelä, S., & Hadwin, A. (2024). Triggers for self-regulated learning: A conceptual framework for advancing multimodal research about SRL. *Learning and Individual Differences*, *115*, 102526.
Järvelä, S., Molenaar, I., & Nguyen, A. (2023). Advancing SRL research with artificial intelligence. *Computers in Human Behavior*, 107847. https://doi.org/10.1016/j.chb.2023.107847
Kahneman, D. (2011). *Thinking, fast and slow*. macmillan.
Kapur, M. (2024). *Productive Failure: Unlocking Deeper Learning Through the Science of Failing*. John Wiley & Sons.
Kasneci, E., Seßler, K., Küchemann, S., Bannert, M., Dementieva, D., Fischer, F., Gasser, U., Groh, G., Günnemann, S., & Hüllermeier, E. (2023). ChatGPT for good? On opportunities and challenges of large language models for education. *Learning and Individual Differences*, *103*, 102274.
Kautz, H. (2022). The third ai summer: Aaai robert s. Engelmore memorial lecture. *Ai Magazine*, *43*(1), 105–125.
Khosravi, H., Shum, S. B., Chen, G., Conati, C., Tsai, Y.-S., Kay, J., Knight, S., Martinez-Maldonado, R., Sadiq, S., & Gašević, D. (2022). Explainable Artificial Intelligence in education. *Computers and Education: Artificial Intelligence*, *3*, 100074. https://doi.org/10.1016/j.caeai.2022.100074
Kingma, D. P., & Welling, M. (2019). An introduction to variational autoencoders. *Foundations and Trends® in Machine Learning*, *12*(4), 307–392.
Kitto, K., Hicks, B., & Buckingham Shum, S. (2023). Using causal models to bridge the divide between big data and educational theory. *British Journal of Educational Technology*, *54*(5), 1095–1124. https://doi.org/10.1111/bjet.13321
Knight, S., Wise, A. F., & Chen, B. (2017). Time for change: Why learning analytics needs temporal analysis. *Journal of Learning Analytics*, *4*(3), 7–17. https://doi.org/10.18608/jla.2017.43.2
Kohavi, R., & John, G. H. (1997). Wrappers for feature subset selection. *Artificial Intelligence*, *97*(1–2), 273–324.
Koller, D., & Sahami, M. (1996). *Toward optimal feature selection*. Stanford InfoLab.
Kononenko, I., Šimec, E., & Robnik-Šikonja, M. (1997). Overcoming the myopia of inductive learning algorithms with RELIEFF. *Applied Intelligence*, *7*, 39–55.
Kotu, V., & Deshpande, B. (2014). *Predictive analytics and data mining: Concepts and practice with rapidminer*. Morgan Kaufmann. https://doi.org/10.1016/C2014-0-00329-2

Kovacic, Z. (2010). *Early prediction of student success: Mining students' enrolment data*.
Krishna, S., Han, T., Gu, A., Pombra, J., Jabbari, S., Wu, S., & Lakkaraju, H. (2022). *The Disagreement Problem in Explainable Machine Learning: A Practitioner's Perspective* (arXiv:2202.01602). arXiv. http://arxiv.org/abs/2202.01602
Krizhevsky, A., Sutskever, I., & Hinton, G. E. (2012). Imagenet classification with deep convolutional neural networks. *Advances in Neural Information Processing Systems*, *25*.
Kumar, H., Musabirov, I., Reza, M., Shi, J., Wang, X., Williams, J. J., Kuzminykh, A., & Liut, M. (2023). Impact of guidance and interaction strategies for LLM use on Learner Performance and perception. *arXiv Preprint arXiv:2310.13712*.
Kumar, I. E., Venkatasubramanian, S., Scheidegger, C., & Friedler, S. (2020). *Problems with Shapley-value-based explanations as feature importance measures*. 5491–5500.
Lakkaraju, H., & Bastani, O. (2020). *"How do I fool you?" Manipulating User Trust via Misleading Black Box Explanations*. 79–85. https://doi.org/10.48550/arXiv.1911.06473
Lample, G., & Charton, F. (2019). Deep learning for symbolic mathematics. *arXiv Preprint arXiv:1912.01412*.
Lang, C., Siemens, G., Wise, A., Gašević, D., & Merceron, A. (Eds.). (2022). *Handbook of learning analytics* (Second). Society for Learning Analytics and Research.
Latif, E., & Zhai, X. (2024). Fine-tuning ChatGPT for automatic scoring. *Computers and Education: Artificial Intelligence*, *6*, 100210.
Laugel, T., Renard, X., Lesot, M.-J., Marsala, C., & Detyniecki, M. (2018). Defining locality for surrogates in post-hoc interpretablity. *arXiv Preprint arXiv:1806.07498*.
Lavelle-Hill, R., Frenzel, A. C., Goetz, T., Lichtenfeld, S., Marsh, H. W., Pekrun, R., Sakaki, M., Smith, G., & Murayama, K. (2024). How the predictors of math achievement change over time: A longitudinal machine learning approach. *Journal of Educational Psychology*.
LeCun, Y., & Bengio, Y. (1995). Convolutional networks for images, speech, and time series. *The Handbook of Brain Theory and Neural Networks*, *3361*(10), 1995.
Lee, J., Hicke, Y., Yu, R., Brooks, C., & Kizilcec, R. F. (2024). The life cycle of large language models in education: A framework for understanding sources of bias. *British Journal of Educational Technology*, *55*(5), 1982–2002.
Lee, S., & Chung, J. Y. (2019). The machine learning-based dropout early warning system for improving the performance of dropout prediction. *Applied Sciences*, *9*(15), 3093.
Lenat, D. B., Prakash, M., & Shepherd, M. (1985). CYC: Using common sense knowledge to overcome brittleness and knowledge acquisition bottlenecks. *AI Magazine*, *6*(4), 65–65.
Li, L., Sha, L., Li, Y., Raković, M., Rong, J., Joksimovic, S., Selwyn, N., Gašević, D., & Chen, G. (2023). Moral machines or tyranny of the majority? A systematic review on predictive bias in education. *Proceedings of the 13th International Learning Analytics and Knowledge Conference*, 499–508.
Li, L., Srivastava, N., Rong, J., Guan, Q., Gašević, D., & Chen, G. (2025). When and how biases seep in: Enhancing debiasing approaches for fair educational predictive analytics. *British Journal of Educational Technology*, in press. https://doi.org/10.1111/bjet.13575
Li, L., & Wang, Z. (2023). Calibrated q-matrix-enhanced deep knowledge tracing with relational attention mechanism. *Applied Sciences*, *13*(4), 2541.
Lo, C. K. (2023). What is the impact of ChatGPT on education? A rapid review of the literature. *Education Sciences*, *13*(4), 410.

Lo Piano, S. (2020). Ethical principles in machine learning and artificial intelligence: Cases from the field and possible ways forward. *Humanities and Social Sciences Communications*, *7*(1), 1–7.
Lundberg, S. M., & Lee, S.-I. (2017). A unified approach to interpreting model predictions. *Advances in Neural Information Processing Systems*, *30*.
Lyu, W., Wang, Y., Chung, T., Sun, Y., & Zhang, Y. (2024). *Evaluating the effectiveness of llms in introductory computer science education: A semester-long field study*. 63–74.
Mao, J., Gan, C., Kohli, P., Tenenbaum, J. B., & Wu, J. (2019). The neuro-symbolic concept learner: Interpreting scenes, words, and sentences from natural supervision. *arXiv Preprint arXiv:1904.12584*.
Mao, Y. (2018). Deep Learning vs. Bayesian Knowledge Tracing: Student Models for Interventions. *Journal of Educational Data Mining*, *10*(2).
Maree, C., Modal, J. E., & Omlin, C. W. (2020). *Towards responsible AI for financial transactions*. 16–21.
Maron, O., & Moore, A. (1993). Hoeffding races: Accelerating model selection search for classification and function approximation. *Advances in Neural Information Processing Systems*, *6*.
McDermott, J. (1982). R1: A rule-based configurer of computer systems. *Artificial Intelligence*, *19*(1), 39–88.
Miller, T. (2019). Explanation in artificial intelligence: Insights from the social sciences. *Artificial Intelligence*, *267*, 1–38.
Miroyan, M., Mitra, C., Jain, R., Ranade, G., & Norouzi, N. (2025). *Analyzing Pedagogical Quality and Efficiency of LLM Responses with TA Feedback to Live Student Questions*. 770–776.
Molenaar, I., Mooij, S. de, Azevedo, R., Bannert, M., Järvelä, S., & Gašević, D. (2023). Measuring self-regulated learning and the role of AI: Five years of research using multimodal multichannel data. *Computers in Human Behavior*, *139*, 107540. https://doi.org/10.1016/j.chb.2022.107540
Molnar, C. (2020). *Interpretable machine learning*. Lulu. com.
Mouta, A., Pinto-Llorente, A. M., & Torrecilla-Sánchez, E. M. (2024). Uncovering blind spots in education ethics: Insights from a systematic literature review on artificial intelligence in education. *International Journal of Artificial Intelligence in Education*, *34*(3), 1166–1205.
Murdoch, W. J., Singh, C., Kumbier, K., Abbasi-Asl, R., & Yu, B. (2019). Definitions, methods, and applications in interpretable machine learning. *Proceedings of the National Academy of Sciences*, *116*(44), 22071–22080.
Nath, D., Gasevic, D., Fan, Y., & Rajendran, R. (2024). CTAM4SRL: A Consolidated Temporal Analytic Method for Analysis of Self-Regulated Learning. *Proceedings of the 14th Learning Analytics and Knowledge Conference*, 645–655.
Nazeri, S. (2024). *Exploring temporality in learning analytics: A comprehensive framework for temporal educational studies*.
Page, M. J., McKenzie, J. E., Bossuyt, P. M., Boutron, I., Hoffmann, T. C., Mulrow, C. D., Shamseer, L., Tetzlaff, J. M., Akl, E. A., & Brennan, S. E. (2021). The PRISMA 2020 statement: An updated guideline for reporting systematic reviews. *Bmj*, *372*.
Pan, J., Dong, Z., Yan, L., & Cai, X. (2024). Knowledge graph and personalized answer sequences for programming knowledge tracing. *Applied Sciences*, *14*(17), 7952.

Panadero, E. (2017). A Review of Self-regulated Learning: Six Models and Four Directions for Research. *Frontiers in Psychology*, *8*, Article No. 422. https://doi.org/10.3389/fpsyg.2017.00422

Pargman, T. C., McGrath, C., & Milrad, M. (2024). Towards Responsible AI in Education: Challenges and Implications for Research and Practice. *Computers and Education: Artificial Intelligence*, 100345.

Pearl, J. (1988). *Probabilistic reasoning in intelligent systems: Networks of plausible inference*. Morgan kaufmann.

Pedro, D., & Lowd, D. (2009). Markov Logic: An Interface Layer for Artificial Intelligence. *Morgan & Claypool*.

Pekrun, R. (2006). The control-value theory of achievement emotions: Assumptions, corollaries, and implications for educational research and practice. *Educational Psychology Review*, *18*, 315–341. https://doi.org/10.1007/s10648-006-9029-9

Pelánek, R. (2017). Bayesian knowledge tracing, logistic models, and beyond: An overview of learner modeling techniques. *User Modeling and User-Adapted Interaction*, *27*, 313–350. https://doi.org/10.1007/s11257-017-9193-2

Peña-Ayala, A. (2014). Educational data mining. *Studies in Computational Intelligence*, *524*.

Pennington, J., Socher, R., & Manning, C. D. (2014). *Glove: Global vectors for word representation*. 1532–1543.

Piech, C., Bassen, J., Huang, J., Ganguli, S., Sahami, M., Guibas, L. J., & Sohl-Dickstein, J. (2015). Deep knowledge tracing. *Advances in Neural Information Processing Systems*, *28*. https://doi.org/10.48550/arXiv.1506.05908

Platzer, A. (2024). *Intersymbolic AI: Interlinking symbolic AI and subsymbolic AI*. 162–180.

Porayska-Pomsta, K. (2024). A manifesto for a pro-actively responsible AI in education. *International Journal of Artificial Intelligence in Education*, *34*(1), 73–83.

Porayska-Pomsta, K., Holmes, W., & Nemorin, S. (2023). The ethics of AI in education. In *Handbook of Artificial Intelligence in Education* (pp. 571–604). Edward Elgar Publishing. https://doi.org/10.4337/9781800375413.00038

Quinlan, J. R. (2014). *C4. 5: Programs for machine learning*. Elsevier.

Ramaswami, G., Susnjak, T., & Mathrani, A. (2023). Effectiveness of a learning analytics dashboard for increasing student engagement levels. *Journal of Learning Analytics*, *10*(3), 115–134.

Reimann, P. (2009). Time is precious: Variable-and event-centred approaches to process analysis in CSCL research. *International Journal of Computer-Supported Collaborative Learning*, *4*, 239–257.

Resnik, P. (2024). Large language models are biased because they are large language models. *arXiv Preprint arXiv:2406.13138*.

Ribeiro, M. T., Singh, S., & Guestrin, C. (2016). Model-agnostic interpretability of machine learning. *arXiv Preprint arXiv:1606.05386*.

Romero, C., Ventura, S., Pechenizkiy, M., & Baker, R. Sj. (2010). *Handbook of educational data mining*. CRC press.

Rudin, C. (2019). Stop explaining black box machine learning models for high stakes decisions and use interpretable models instead. *Nature Machine Intelligence*, *1*(5), 206–215.

Rumelhart, D. E., Hinton, G. E., & Williams, R. J. (1986). Learning representations by back-propagating errors. *Nature*, *323*(6088), 533–536.

Saarela, M., Heilala, V., Jääskelä, P., Rantakaulio, A., & Kärkkäinen, T. (2021). Explainable student agency analytics. *IEEE Access*, *9*, 137444–137459. https://doi.org/10.1109/access.2021.3116664

Saint, J., Fan, Y., Pardo, A., & Gašević, D. (2022). Temporally-focused analytics of self-regulated learning: A systematic review of literature. *Computers & Education: Artificial Intelligence*, in press.

Sakamoto, Y., Ishiguro, M., & Kitagawa, G. (1986). Akaike information criterion statistics. *Dordrecht, The Netherlands: D. Reidel*, *81*(10.5555), 26853.

Sawyer, R. K. (2022). An introduction to the learning sciences. *The Cambridge Handbook of the Learning Sciences*, 1–24.

Self, J. (1986). The application of machine learning to student modelling. *Instructional Science*, *14*(3), 327–338.

Selvaraju, R. R., Cogswell, M., Das, A., Vedantam, R., Parikh, D., & Batra, D. (2017). *Grad-cam: Visual explanations from deep networks via gradient-based localization*. 618–626.

Serafini, L., Donadello, I., & Garcez, A. d'Avila. (2017). *Learning and reasoning in logic tensor networks: Theory and application to semantic image interpretation*. 125–130.

Sha, L., Li, Y., Gasevic, D., & Chen, G. (2022). Bigger data or fairer data?: Augmenting BERT via active sampling for educational text classification. *International Conference on Computational Linguistics 2022*, 1275–1285.

Shakya, A., Rus, V., & Venugopal, D. (2021). Student Strategy Prediction Using a Neuro-Symbolic Approach. *International Educational Data Mining Society*.

Siemens, G. (2013). Learning analytics: The emergence of a discipline. *American Behavioral Scientist*, *57*(10), 1380–1400.

Silver, D., Huang, A., Maddison, C. J., Guez, A., Sifre, L., Van Den Driessche, G., Schrittwieser, J., Antonoglou, I., Panneershelvam, V., & Lanctot, M. (2016). Mastering the game of Go with deep neural networks and tree search. *Nature*, *529*(7587), 484–489.

Singh, C., Inala, J. P., Galley, M., Caruana, R., & Gao, J. (2024). Rethinking interpretability in the era of large language models. *arXiv Preprint arXiv:2402.01761*.

Slack, D., Hilgard, S., Jia, E., Singh, S., & Lakkaraju, H. (2020). Fooling LIME and SHAP: Adversarial Attacks on Post hoc Explanation Methods. *Proceedings of the AAAI/ACM Conference on AI, Ethics, and Society*, 180–186. https://doi.org/10.1145/3375627.3375830

Sourek, G., Aschenbrenner, V., Zelezny, F., Schockaert, S., & Kuzelka, O. (2018). Lifted relational neural networks: Efficient learning of latent relational structures. *Journal of Artificial Intelligence Research*, *62*, 69–100. https://doi.org/10.1613/jair.1.11203

Sperling, K., Stenliden, L., Nissen, J., & Heintz, F. (2024). Behind the scenes of co-designing AI and LA in K-12 education. *Postdigital Science and Education*, *6*(1), 321–341.

Stevens, R., Taylor, V., Nichols, J., Maccabe, A. B., Yelick, K., & Brown, D. (2020). *Ai for science: Report on the department of energy (doe) town halls on artificial intelligence (ai) for science*. Argonne National Lab.(ANL), Argonne, IL (United States). https://doi.org/10.2172/1604756

Sutton, C., & McCallum, A. (2012). An introduction to conditional random fields. *Foundations and Trends® in Machine Learning*, *4*(4), 267–373.

Svatos, M., Sourek, G., & Zelezný, F. (2019). *Revisiting Neural-Symbolic Learning Cycle*. NeSy@ IJCAI.

Sweller, J. (2023). The development of cognitive load theory: Replication crises and incorporation of other theories can lead to theory expansion. *Educational Psychology Review*, *35*(4), 95.

Taiye, M., High, C., Velander, J., Matar, K., Okmanis, R., & Milrad, M. (2024). *Generative ai-enhanced academic writing: A stakeholder-centric approach for the design and development of chat4isp-ai*. 74–80.

Tato, A., & Nkambou, R. (2022). Infusing expert knowledge into a deep neural network using attention mechanism for personalized learning environments. *Frontiers in Artificial Intelligence*, *5*, 921476.

Toomla, K., Xiaojing, W., Kikas, E., Malleus-Kotšegarov, E., Aus, K., Azevedo, R., & Hooshyar, D. (2025). Measuring (meta)emotion, (meta)motivation, and (meta)cognition using digital trace data: A systematic review of K-12 self-regulated learning. *Proceedings of the 40th ACM/SIGAPP Symposium on Applied Computing*. ACM SAC, Italy. https://doi.org/10.1145/3672608.3707961

Torralba, A., & Efros, A. A. (2011). *Unbiased look at dataset bias*. 1521–1528. https://doi.org/10.1109/CVPR.2011.5995347

Türker, M. A., & Zingel, S. (2008). Formative interfaces for scaffolding self-regulated learning in PLEs. *Elearning Papers*, *14*(9). https://doi.org/10.3390/app10062145

UNESCO. (2019). *Beijing consensus on artificial intelligence and education*.

Valero-Leal, E., Carlon, M. K. J., & Cross, J. S. (2023). *A shap-inspired method for computing interaction contribution in deep knowledge tracing*. 460–465.

Van den Broeck, G., Lykov, A., Schleich, M., & Suciu, D. (2022). On the tractability of SHAP explanations. *Journal of Artificial Intelligence Research*, *74*, 851–886.

Van Petegem, C., Deconinck, L., Mourisse, D., Maertens, R., Strijbol, N., Dhoedt, B., De Wever, B., Dawyndt, P., & Mesuere, B. (2023). Pass/fail prediction in programming courses. *Journal of Educational Computing Research*, *61*(1), 68–95.

Venugopal, D., Rus, V., & Shakya, A. (2021). *Neuro-symbolic models: A scalable, explainable framework for strategy discovery from big edu-data*. Proceedings of the 2nd Learner Data Institute Workshop in Conjunction with The 14th International Educational Data Mining Conference. https://ceur-ws.org/Vol-3051/LDI_4.pdf

Vincent-Lancrin, S., & Van der Vlies, R. (2020). Trustworthy artificial intelligence (AI) in education: Promises and challenges. *OECD Education Working Papers*, *218*, 0_1-17.

Waheed, H., Hassan, S.-U., Nawaz, R., Aljohani, N. R., Chen, G., & Gasevic, D. (2023). Early prediction of learners at risk in self-paced education: A neural network approach. *Expert Systems with Applications*, *213*, 118868. https://doi.org/10.1016/j.eswa.2022.118868

Wang, S., Wang, F., Zhu, Z., Wang, J., Tran, T., & Du, Z. (2024). Artificial intelligence in education: A systematic literature review. *Expert Systems with Applications*, *252*, 124167.

Warr, M., Oster, N. J., & Isaac, R. (2024). Implicit bias in large language models: Experimental proof and implications for education. *Journal of Research on Technology in Education*, 1–24.

Webb, G. I. (1989). A machine learning approach to student modelling. *Proceedings of the Third Australian Joint Conference on Artificial Intelligence*, 195–205.

Webb, G. I. (1991). Inside the Unification Tutor: The architecture of an intelligent educational system. *Proceedings of the Fourth Australian Society for Computers in Learning in Tertiary Education Conference*, 677–684.

Weidlich, J., Gašević, D., & Drachsler, H. (2022). Causal Inference and Bias in Learning Analytics: A Primer on Pitfalls Using Directed Acyclic Graphs. *Journal of Learning Analytics*, *9*(3), Article 3. https://doi.org/10.18608/jla.2022.7577

Weigend, A., Rumelhart, D., & Huberman, B. (1990). Generalization by weight-elimination with application to forecasting. *Advances in Neural Information Processing Systems*, *3*.

Weiss, S. M., & Kulikowski, C. A. (1991). *Computer systems that learn: Classification and prediction methods from statistics, neural nets, machine learning, and expert systems*. Morgan Kaufmann Publishers Inc.

Werder, K., Ramesh, B., & Zhang, R. (2022). Establishing data provenance for responsible artificial intelligence systems. *ACM Transactions on Management Information Systems (TMIS)*, *13*(2), 1–23.

White, A., & d'Avila Garcez, A. (2020). Measurable counterfactual local explanations for any classifier. In *ECAI 2020* (pp. 2529–2535). IOS Press.

Winne, P. H. (2018). Theorizing and researching levels of processing in self-regulated learning. *British Journal of Educational Psychology*, *88*(1), 9–20.

Xu, H., Yin, J., Qi, C., Gu, X., Jiang, B., & Zheng, L. (2024). Bridging the Vocabulary Gap: Using Side Information for Deep Knowledge Tracing. *Applied Sciences*, *14*(19), 8927.

Yan, L., Greiff, S., Teuber, Z., & Gašević, D. (2024). Promises and challenges of generative artificial intelligence for human learning. *Nature Human Behaviour*, *8*(10), 1839–1850. https://doi.org/10.1038/s41562-024-02004-5

Yan, L., Sha, L., Zhao, L., Li, Y., Martinez-Maldonado, R., Chen, G., Li, X., Jin, Y., & Gašević, D. (2024). Practical and ethical challenges of large language models in education: A systematic scoping review. *British Journal of Educational Technology*, *55*(1), 90–112.

Yang, Y., Hooshyar, D., Pedaste, M., Wang, M., Huang, Y.-M., & Lim, H. (2020). Predicting course achievement of university students based on their procrastination behaviour on Moodle. *Soft Computing*, *24*, 18777–18793.

Yeung, C.-K., & Yeung, D.-Y. (2018). *Addressing two problems in deep knowledge tracing via prediction-consistent regularization*. 1–10.

Yun, H., Song, H.-D., & Kim, Y. (2025). Identifying university students' online self-regulated learning profiles: Predictors, outcomes, and differentiated instructional strategies. *European Journal of Psychology of Education*, *40*(1), 1–25.

Zhao, W., Xia, J., Jiang, X., & He, T. (2023). A novel framework for deep knowledge tracing via gating-controlled forgetting and learning mechanisms. *Information Processing & Management*, *60*(1), 103114.

Zhidkikh, D., Heilala, V., Van Petegem, C., Dawyndt, P., Jarvinen, M., Viitanen, S., De Wever, B., Mesuere, B., Lappalainen, V., & Kettunen, L. (2024). Reproducing Predictive Learning Analytics in CS1: Toward Generalizable and Explainable Models for Enhancing Student Retention. *Journal of Learning Analytics*, *11*(1), 132–150.